**Enhanced Tantalum Superconducting Resonator Performance via All-Surface Organic Monolayer Passivation**

*Harsh Gupta*[#*], *Moritz Singer*[#], *Benedikt Schoof, Anna Cattani-Scholz, Shreya Sharma, Luca Rommeis, Marc Tornow*[*]

Harsh Gupta, Moritz Singer, Benedikt Schoof, Anna Cattani-Scholz, Luca Rommeis, Marc Tornow
*School of Computation, Information and Technology, Department of Electrical Engineering, Technical University of Munich, 85748 Garching, Germany*
E-Mail*: harsh.gupta@tum.de, tornow@tum.de

Shreya Sharma
*Department of Biosciences and Bioengineering, Indian Institute of Technology Roorkee, 247667, Roorkee, India*

Luca Rommeis
*Fraunhofer Institute for Electronic Microsystems and Solid State Technologies (EMFT), 80686 Munich, Germany*

Funding: We acknowledge funding through the Munich Quantum Valley project by the Free State of Bavaria, Germany and the Munich Quantum Valley Quantum Computer Demonstrators- Superconducting Qubits (MUNIQC-SC) project (grant no. 13N16188), funded by the Federal Ministry of Research, Technology and Space, Germany

# Authors contributed equally to this work

Tantalum is a promising platform for superconducting quantum circuits, yet coherence times remain limited by dielectric losses from interfacial two-level systems (TLS), exacerbated by native oxide regrowth. Here, we implement molecular surface passivation using self-assembled organic monolayers on freshly etched tantalum and silicon in coplanar waveguide resonators. Surface characterization by contact angle, XPS, FTIR and TEM confirms the formation of ordered, nanometer-thick films that suppress oxide formation. Microwave measurements in the ~5-9 GHz range reveal internal quality factors up to $1.8\times10^6$ in the single-photon regime at 100 mK, representing a ~140% improvement over untreated devices with native oxide. Power and temperature dependent measurements attribute this enhancement to reduced TLS-induced losses. These results demonstrate that molecular passivation effectively engineers low-loss interfaces and provides a scalable route toward high-coherence superconducting quantum devices.

## 1. Introduction

Superconducting quantum technologies, which primarily utilize Josephson junctions (JJs) and coplanar waveguide (CPW) microwave resonators, are promising in numerous applications such as quantum memory,[1–3] sensing[4–7] and superconducting qubits.[8–13] For superconducting qubits, artificial atoms can be realized with JJs coupled to planar capacitances and microwave resonators, which serve for qubit state manipulation and readout.[14–19] In order to develop practical qubits, the qubit states need to have the longest possible coherence times, a critical factor affected and limited by various losses, primarily two-level system (TLS) and quasiparticle (QP) losses.[19–23] Superconducting CPW resonators can be a reliable tool for characterizing and understanding different losses in superconducting quantum circuits.[20,24–27] The behavior of these losses can be quantified by the internal quality factor ($Q_i$) of these microwave resonators and their corresponding resonant frequencies, when characterized at various input powers and temperatures.

TLS losses in quantum circuits can arise from defects, tunnelling atoms, dangling bonds, hydrogen defects or collective motions of small atomic groups; they are primarily hosted by polar impurities (like -OH) and/or oxygen bridges in dielectric native oxides formed on superconductor metal-to-air and substrate-to-air interfaces upon air exposure.[20,28] TLS losses are significantly dominating and are the limiting factor in the low power and low temperature regime of operation for qubits.[20,22] In contrast, quasiparticles (QPs) are intrinsic to the superconductor itself, either thermally excited with increase in temperature or generated as non-equilibrium QPs by stray radiation or input microwave power.[29–33] As the QP density increases, the change in surface impedance affects the resonant frequency and reduces the $Q_i$ of the resonators.[29,33–35] From a materials aspect, a continuous effort has been made to explore different superconductors such as aluminum, niobium, titanium nitride, rhenium etc. to improve the performance of quantum circuits. [36,37] However, these materials present a challenge owing to high dielectric losses from their native oxides at the metal-air interface.[36,38] This has recently led to an increased interest in tantalum, especially in its α-phase, as promising alternative superconducting material for quantum circuits, with reported milliseconds coherence times.[39–44] It is reported that, tantalum has fewer sub-oxides and a smoother interface than niobium and its oxide; further, tantalum oxide favors surfaces with fewer oxygen dangling bonds compared to niobium oxide.[45–47] These characteristics are held responsible for lower measured dielectric TLS losses in tantalum and hence improved performance. Even though tantalum offers better material aspects than niobium, it is still limited by native oxide formation at the metal-air and substrate-air interfaces. In the case of tantalum, the native $TaO_x$ is commonly removed by selective wet etching, e.g., in Buffered-Oxide Etch (BOE) solution, but this does not restrict the subsequent re-growth of native oxide. To address this challenge, one can suppress the regrowth of native oxides at the various interfaces by capping the freshly etched surfaces with a suitable passivation layer. For example, recently, the improvement of Nb qubits' coherence times by encapsulating the structures with Ta or Au metallic thin-films has been reported.[48,49] In the case of tantalum, Chang *et al.* reported noble metal encapsulation of resonators leading to suppressed native oxide growth,[50] which however was limited to the planar, horizontal Ta surface plane on top of the resonators. The vertical edges of the devices could not be covered/passivated using the employed resonator fabrication and metal encapsulation process, resulting in a still limited device performance.

This problem calls for a passivation process which can cap all exposed resonator surfaces, including the edges. A facile and convenient methodology is using solution-based growth of organic molecule self-assembled monolayers (SAMs), which would readily passivate all accessible surfaces (substrate-air and metal-air interface) - both all around the entire resonator structure and on the open substrate surfaces nearby. SAMs consist of a single molecular layer, a few nanometers thick, composed of short organic molecules organized in a well-defined, closely packed, 2D crystalline-like structure. Their formation is driven by the intrinsic self-assembly behavior of these molecules, primarily governed by van der Waals and hydrophobic interactions, hydrogen bonding and chemisorption at the substrate surface.[51,52] Well established classes of SAM-forming molecules, distinguished by their surface-binding headgroups, include alkenes, thiols, silanes, carboxylic acids and phosphonic acids.[51–54] They can tailor numerous properties such as wettability, work function and chemical reactivity of different materials. This led to the application of SAMs in diverse fields including surface passivation, nanotechnology, electronic devices and biosensors.[55–59] Recently, Alghadeer *et al.* have reported the effect of passivation of niobium resonators using octatrichlorosilane molecules,[60] also we have demonstrated that a high temporal stability of Nb resonators can be achieved by organophosphonate SAM passivation.[27] Despite a few reports on the SAM-based functionalization of tantalum oxide,[61–64] the broader impact of organic SAM coatings on conventional elemental superconductors, here in particular on tantalum, towards quantum technologies has remained largely unexplored. Moreover, realizing single molecule SAM growth on both superconducting tantalum films and silicon substrates for their simultaneous passivation poses a significant materials integration challenge.

In this work, we investigate the influence of alkene-based SAM passivation on the performance of α-tantalum CPW microwave resonators. Alkene SAMs are employed for the growth of well-ordered molecular layers that simultaneously passivate both the metal-air (tantalum-air) and substrate-air (silicon-air) interfaces, thereby inhibiting the re-growth of native oxides. We systematically analyze the internal quality factors, loss mechanisms and resonant frequency shifts of the resonators as a function of input power and temperature. Comparative studies are performed between resonators featuring this alkene SAM passivation and two types of reference resonators: those covered by their native oxide and those after oxide removal but without subsequent passivation. The structural and chemical modifications induced by SAM formation are further corroborated using surface analytical techniques, including contact angle measurements, x-ray photoelectron spectroscopy (XPS), Fourier-transform infrared (FTIR) spectroscopy and cross-sectional TEM imaging. We show that organic self-assembled monolayers can simultaneously passivate the Si-air and Ta-air interfaces, thereby significantly reducing TLS losses and enhancing quality factors of Ta resonators in the single-photon regime. We anticipate this molecular-level approach to offer a scalable route to the interface engineering of next-generation superconducting quantum devices based on tantalum.

## 2. Results and discussion

### 2.1. SAM growth and material characterizations

We deposited ~200 nm thick α-tantalum thin films on high-resistivity Si (100) at room temperature, mediated by a ~5nm niobium seed layer below the tantalum film. Grazing angle incident-X-ray diffraction (GI-XRD) spectra (Figure S1a) clearly show the dominance of the α-phase within the Ta thin films. Measurement of the Ta film's critical temperature and residual resistance ratio (RRR = $R_{270K}/R_{5K}$) yields a $T_c$ of ~4.2 K (Figure S1b) and an RRR of approximately 2, respectively, which agrees with reported values for α-tantalum thin films.[39,42,65] In the following, and throughout this work, samples that did not undergo BOE treatment following fabrication are designated as '*native oxide*', those treated exclusively with BOE as '*unpassivated*', and those subjected to SAM coating after BOE treatment as '*passivated*'.

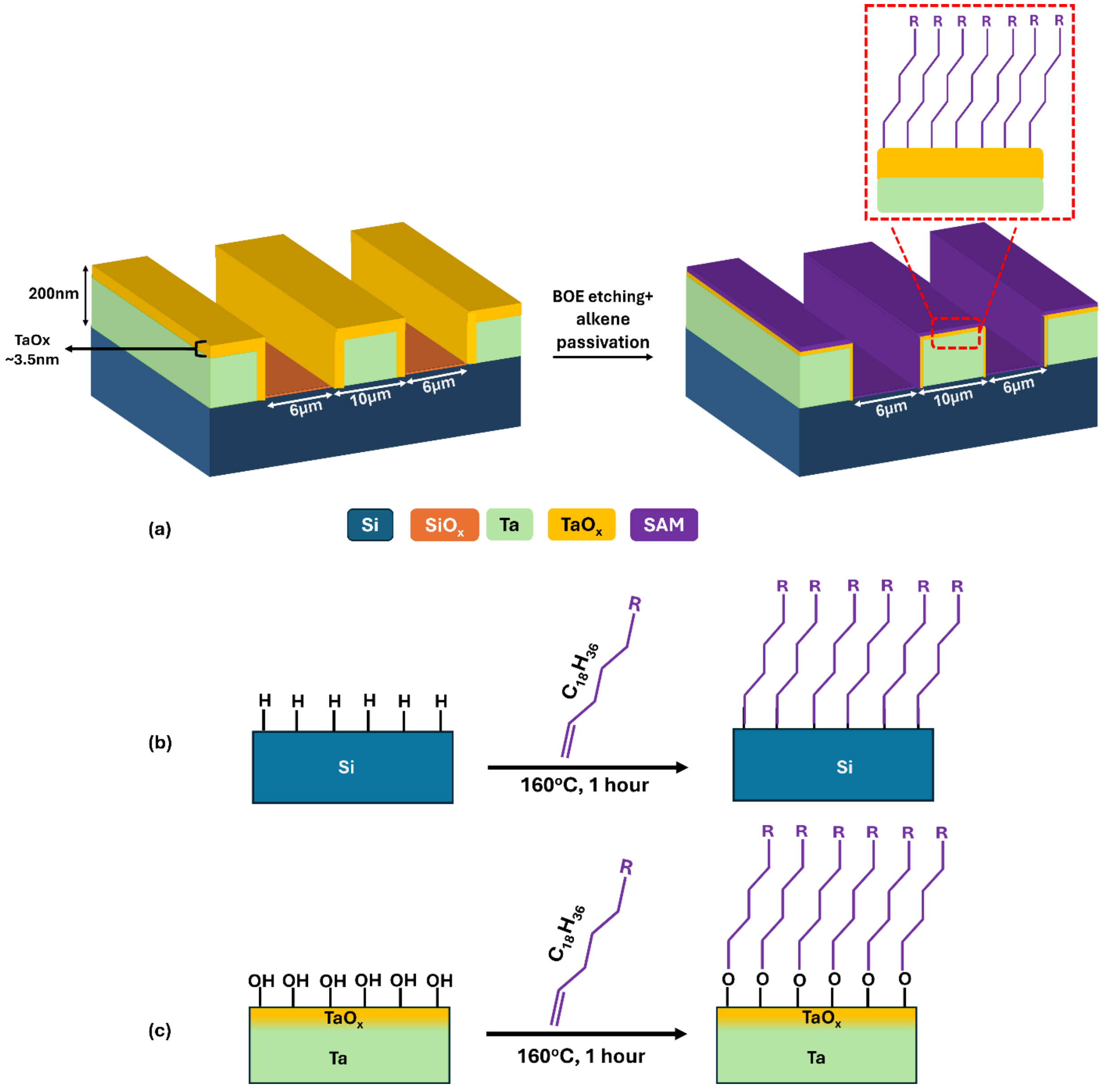

***Figure 1***. *(a) Schematic showing the cross-section of a resonator comprising α-Ta thin film of thickness 200 nm, covered by a $TaO_x$ (native oxide) layer of estimated thickness ~3.5nm, before (left) and after (right) passivation of the two different, exposed surfaces (note that the thin Nb seed layer below Ta is not included in the image); (b) Growth mechanism for an alkene SAM*

*on H-terminated Si (hydrosilylation), mediated by thermal activation of the surface; (c) Proposed growth mechanism on hydroxylated (-OH terminated) Ta native oxide.*

Self-assembled monolayers (SAMs) were prepared on Si and Ta substrates following BOE etching. On Si, this step removes the native oxide and yields an H-terminated surface. Although the substrates were briefly exposed to ambient air prior to transfer into the reaction chamber for SAM growth, hydrogen-terminated Si is known to exhibit resistance to oxidation over short air exposure times, preserving its surface reactivity upon activation. Alkene attachment on Si then proceeds via thermally induced desorption of surface hydrogen atoms by homolytic bond cleavage, creating Si dangling bonds that act as reactive sites for covalent grafting of alkene molecules (**Figure 1b**).[53,66] In contrast, brief air exposure of Ta would result in the spontaneous re-formation of a thin, presumably non-stoichiometric native oxide layer. This oxide layer, however, can be exploited advantageously, as its unsaturated nature facilitates surface hydroxylation, which likewise provides reactive sites for alkene binding upon thermal activation (Figure 1c). Such thermally assisted alkene grafting mechanisms on hydroxylated metal oxide surfaces have been widely reported in the literature.[53,64,67]

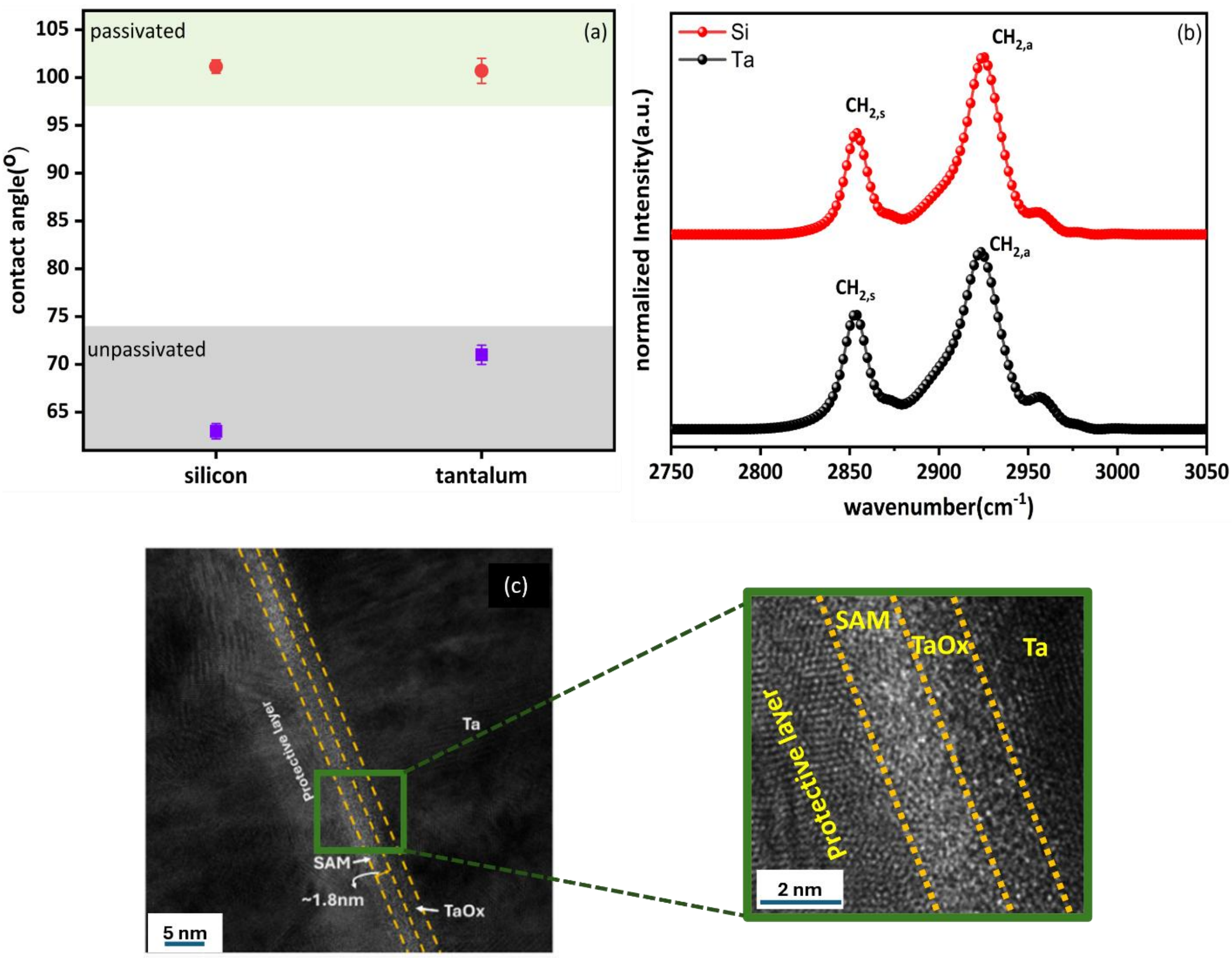

***Figure 2***. *(a) Contact angle values for passivated and unpassivated Si and Ta surfaces; a significant increase in hydrophobicity post passivation is evident. (b) background subtracted FTIR spectra for passivated Si and Ta showing the presence of different vibrational modes of alkyl molecules on the surface. Labels at the peaks indicate: the symmetric stretch vibration ($CH_{2,s}$) and the asymmetric stretch vibration ($CH_{2,a}$) of the methylene-groups of the molecules.*

*(c) Cross-sectional TEM image with close-up on the right, of a passivated Ta thin film; a SAM of thickness ~1.8 nm, sandwiched between the thin native oxide and a protective layer used for FIB preparation can be discerned.*

To assess the structural integrity and quality of the SAMs, alkene monolayers were first grown on planar silicon and tantalum substrates. Subsequently, AFM images were recorded in tapping mode to determine the surface morphology of the SAMs. Specifically, the surface roughness values for passivated Si and Ta films were estimated: we measured root mean squared (RMS) roughness values for passivated Si and Ta thin films of 0.7±0.1 nm and 0.9±0.2 nm, respectively (see, Figure S2). This confirms the uniform and conformal growth of alkene SAMs on Si and Ta. Water contact angle measurements were equally performed on both substrate types to probe changes in surface wettability following alkene SAM deposition. Both surfaces exhibited contact angles ~100° post SAM growth, signifying a transition from hydrophilic to distinct hydrophobic behavior as presented in **Figure 2a**. The pronounced hydrophobicity suggests the formation of densely packed, well-ordered aliphatic SAMs terminated with methyl groups. The high contact angle observed for SAM-passivated Ta and Si surfaces further suggests enhanced resistance to humidity-induced surface oxidation. FTIR spectra of SAM-passivated silicon and tantalum thin films exhibit characteristic absorption bands in the 2750-3050 $cm^{-1}$ range, corresponding to C-H stretching vibrations of aliphatic hydrocarbons (Figure 2b). The symmetric and asymmetric $CH_2$ stretching modes appear at 2854 $cm^{-1}$ and 2923 $cm^{-1}$ for tantalum and at 2854 $cm^{-1}$ and 2924 $cm^{-1}$ for silicon, respectively, consistent with well-ordered and uniform octadecene SAMs on both surfaces [53,68]. To estimate the thickness of our SAMs grown on planar Si and Ta thin films ellipsometry measurements were conducted. The SAM layers were fitted using Cauchy's model with a refractive index value of 1.5,[69] which yielded a measured film thickness of ~2 nm. The presence of the SAM layer on the native tantalum oxide surface was further confirmed by cross-sectional TEM imaging (Figure 2c), which revealed a uniform organic layer with an average thickness of approximately 1.8 nm on tantalum, which is in close agreement with ellipsometry. Complementary cross-sectional electron energy loss spectroscopy (EELS) elemental mapping corroborated these observations, showing a pronounced carbon signal localized at the SAM-oxide interface (Figure S4), consistent with the presence of the alkene monolayer, along with signals from the underlying elemental constituents.

To independently assess the effect of passivation on tantalum thin films, the surface chemical composition was examined using X-ray photoelectron spectroscopy (XPS), as summarized in **Figure 3**. Unpassivated, passivated, and native-oxide Ta thin films were analyzed, with elemental constituents identified from survey spectra. Narrow-scan Ta 4f spectra were subsequently acquired in the 18-34 eV range for all samples. Upon Shirley background correction and peak fitting, two symmetric doublets were attributed to oxidized tantalum species ($Ta_2O_5$ and $TaO_2$), while the asymmetric doublet with a $4f_{7/2}$ component at ~21.8 eV corresponded to metallic Ta. The $4f_{7/2}$ components at ~24.6 eV and ~27.1 eV were previously assigned to $TaO_2$ and $Ta_2O_5$, respectively.[47,70,71] Each oxidation state exhibits the expected spin-orbit splitting characteristic of the Ta 4f level. Notably, the relative Ta signal intensity in the XPS spectra (Figure 3g) is markedly attenuated for the native-oxide sample compared to the passivated film, indicating a thicker oxide layer on the former and a thinner oxide on the latter. The reduced oxide thickness in the passivated sample suggests that the alkene SAM effectively limits surface re-oxidation. Further, we analyze the C-1s spectra for all the samples to probe the presence of alkene molecules on passivated samples. C 1s spectra for passivated

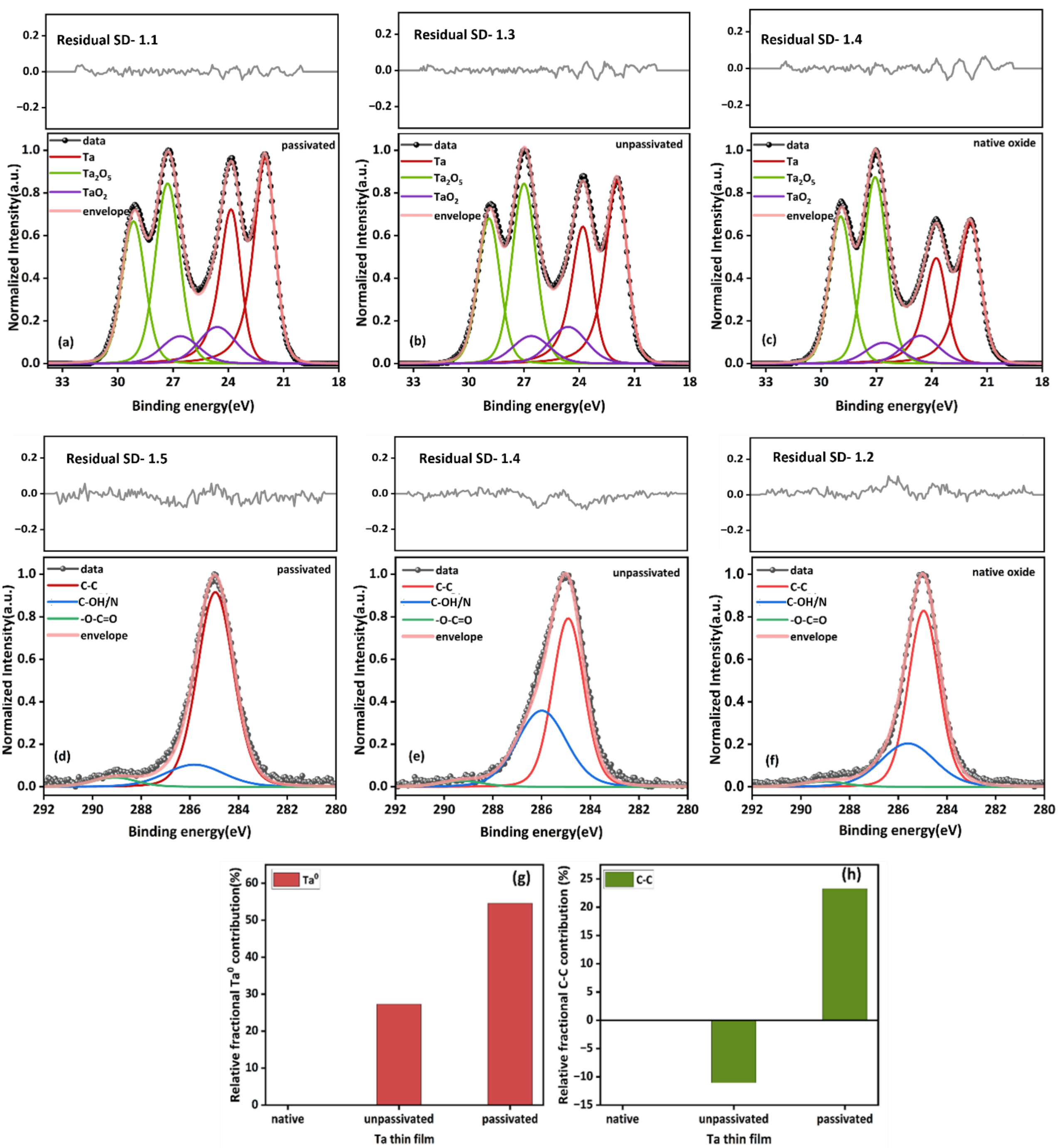

***Figure 3***. *Background corrected and normalized XPS spectra corresponding to Ta 4f and C 1s for (a, d) alkene-passivated Ta (b, e) unpassivated Ta and (c, f) Ta films with native oxide, respectively. The line plots on top of each fitted data represent the residual plots from fitting (difference between raw data and envelope), with extracted standard deviation (SD) for each fit. Different oxidation states for Ta ($Ta^0$: elemental Ta, $TaO_2$ and $Ta_2O_5$) with spin-orbit coupled doublet ($4f_{7/2}$ and $4f_{5/2}$) are observed. Different components corresponding to different oxidation states of C 1s are observed. (g): Relative fractional contribution for $Ta^0$ (elemental tantalum), normalized with respect to native oxide sample (set as zero); passivated films show significantly higher elemental Ta contributions. (h): Relative fractional contribution of C-C*

*bonds; passivated film shows significantly higher C-C contribution, attributed to presence of alkene molecules.*

resonators show significantly higher intensity in comparison to others, indicating the presence of organic layer (Figure S10 b). Each spectrum is deconvoluted with four peaks ascribed to C-C (corresponding to $sp^3$ carbon), C-OH/N and -O-C=O.[72,73] All samples exhibited a prominent contribution from C-C bonded carbon, predominantly attributed to adventitious carbon. Notably, passivated films demonstrated a significantly elevated C-C relative contribution (Figure 3h), which is likely due to the presence of alkene molecules, which comprise a significant portion of $sp^3$-hybridized carbon. For details regarding the fitting routine, relative fractional contribution and thickness estimation see Supporting Information, section “X-ray photoelectron spectroscopy”.

## 2.2 **Microwave measurements**

### *2.2.1 Power dependent $Q_i$ measurements*

Microwave measurements using a Keysight P5002B vector network analyser (VNA) delivered RF signals into the cryostat, where the chip was mounted in a copper box held at 100 mK. The input signal entered the copper box through an SMA connector and reached the chip via a PCB strip line and aluminium bond wires to the centre feedline. The edges of chip were also contacted to the PCB for grounding, and additional bond wires across the feed line were placed to eliminate unwanted electromagnetic field modes. Further, the feedline guided the signal along all twelve capacitively coupled CPW resonators arranged in hanger geometry on both sides of the center feedline (see, Figure 4d), featuring resonant frequencies in the range ~5-9 GHz, before the transmitted output was collected through a second SMA connector and finally transmitted back to the VNA (see also Fig. S5, for a schematic of the complete setup). The internal quality factors ($Q_i$) were subsequently extracted to assess resonator performance (for details, see Supporting Information: Microwave measurement procedure and data analysis), with $Q_i$ inversely related to the internal loss ($\delta_i = 1/Q_i$). **Figure 4a** presents the log-log variation of $Q_i$ as function of mean photon number $\langle n \rangle$, reflecting the input power to the resonator chip. With increasing input power, $Q_i$ rises significantly due to TLS saturation. At higher power ($\langle n \rangle > 10^6$), i.e., beyond the range of our measurements, $Q_i$ is anticipated to approach a plateau, indicating the emergence
of a small, power-independent loss mechanism, such as quasiparticle generation or other, intrinsic losses. This behavior aligns with the understanding that at low powers, TLS-induced losses dominate over quasiparticles and other, power-independent mechanisms.

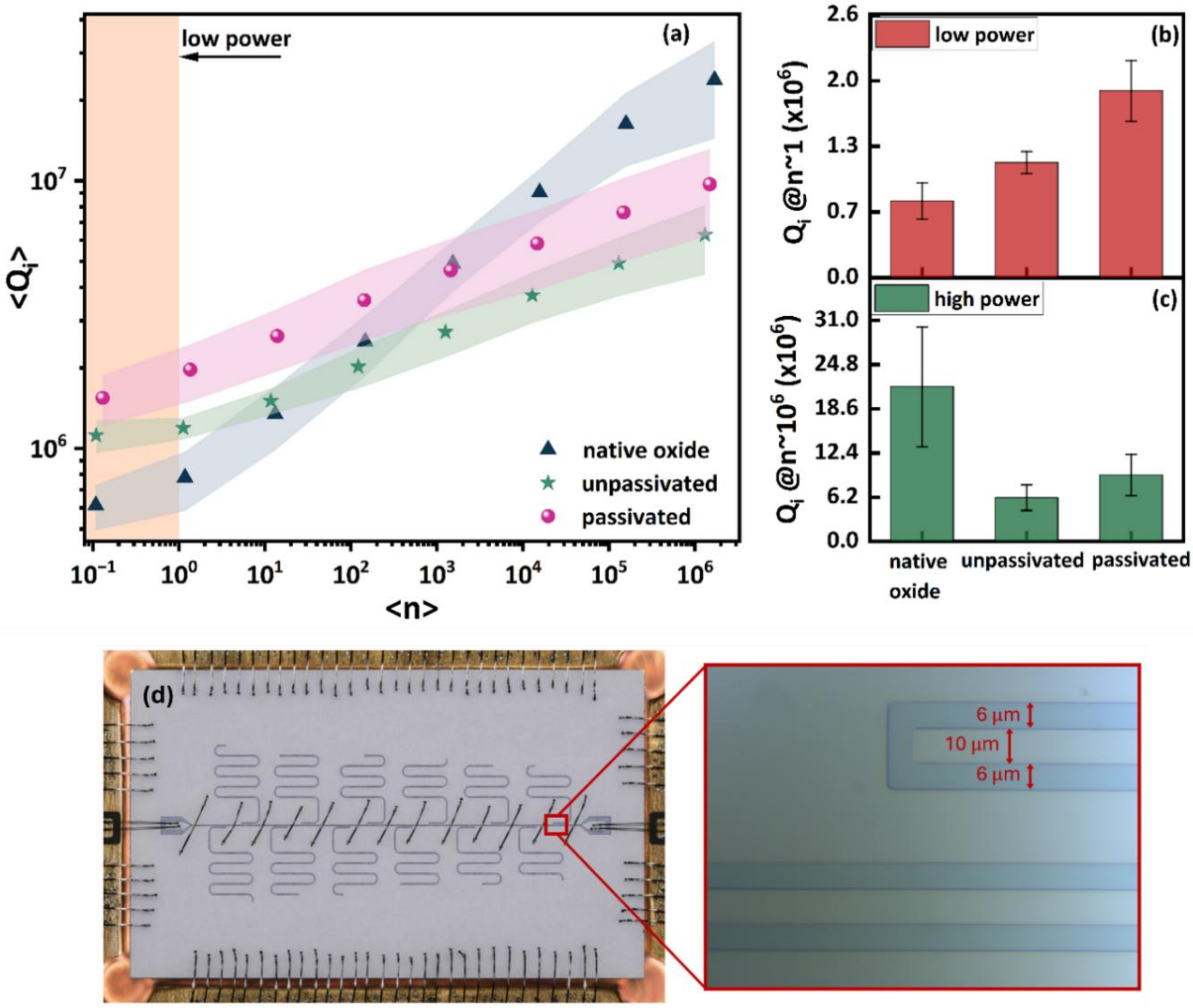

***Figure 4**. (a) Mean internal quality factor ($Q_i$) for Ta resonators at 100 mK against mean photon number, for passivated, unpassivated and resonators with native oxide. The shaded region along the data points in the plot denotes the 95% confidence interval of the data. (b) $Q_i$ at low input power ($\langle n \rangle \sim 1$) and (c) high input power ($\langle n \rangle \sim 10^6$). Distinct behavioral trends of the resonators in the low and high-power regimes can be discerned. Error bars denote the standard deviation from averaging over all resonators. (d) Left: Optical micrograph of a tantalum resonator chip, showing the typical layout design in hanger geometry with resonance frequencies in the range ~5-9 GHz. For the exact distribution of resonant frequencies under study, refer to Fig. S8 (Supporting Information). Wire-bonds surrounding the resonators connect the sample's ground plane to the copper box housing. Right: Close-up, showing the measured dimensions of the centre line and the gaps to the ground plane, 10 µm and 6 µm, respectively.*

In the single-photon regime, tantalum resonators with native oxide exhibit a mean $Q_i$ of $0.78\times10^6$. Following buffered oxide etch (BOE) treatment, $Q_i$ increases to $1.12\times10^6$, and further improves to $1.85\times10^6$ after SAM passivation. This corresponds to an enhancement of approximately 140% compared to the native oxide devices, highlighting the efficacy of molecular passivation in mitigating surface-related losses. Interestingly, in the high-power regime ($\langle n \rangle > 10^6$), resonators with native oxide exhibit the highest $Q_i$, in contrast to their behavior at low powers. This inversion suggests that TLS-related losses are substantially suppressed in the passivated devices, while the residual, power-independent losses are heightened in passivated and unpassivated resonators. We attribute these additional losses to possibly originate in hydrogen related defects post BOE etching, which has been shown to affect the performance of tantalum resonators.[74] At a fixed low temperature and low excitation power, $Q_i$ is primarily limited by TLS losses. The $Q_i$ behavior with input excitation power can be described by equation 1:[75]

$$\frac{1}{Q_i(n)} = \frac{1}{Q_{TLS}^0}\left[\frac{\tanh\left(\frac{\hbar\omega}{2k_BT}\right)}{\sqrt{1+\left(\frac{n}{n_c}\right)^{\beta}}}\right] + \frac{1}{Q_{other}} \qquad ...1$$

where $\hbar$ is the reduced Planck constant, $n$ is the average photon number, $T$ is the temperature, $\omega$ is the angular frequency of the resonator, $Q_{TLS}^0$ is the low-power TLS-limited quality factor (equivalent to the inverse of the characteristic TLS loss), $k_B$ is the Boltzmann constant, $n_c$ is the critical photon number corresponding to the saturation power for TLSs, $\beta$ is an empirical exponent that controls how the TLS loss saturates with photon number, and $Q_{other}$ denotes contributions from other, power-independent losses. We here focus on the extraction of $Q_{TLS}^0$ as it reveals differences in TLS losses in the resonators under different treatments. In principle, $Q_{TLS}^0$ can be obtained by fitting equation 1 to the data, however, as we do not see a clear saturation behavior of $Q_i$ at the highest powers we probe, it becomes difficult to unambiguously distinguish between different loss channels contributing at different powers. Such approach might therefore lead to overestimated values or increased variability of TLS losses as it does not fully account for other loss mechanisms.

Therefore, to identify and interpret the underlying loss mechanisms more clearly, we performed additional temperature-dependent measurements at various input powers, providing a comprehensive database to separate different contributions to the total loss. In superconducting microwave resonators, temperature-dependent shifts in the resonant frequency and changes in internal dissipation arise from both TLS and quasiparticle dynamics. TLS in dielectrics or at interfaces modify the complex dielectric function, with $Re[\varepsilon_{TLS}(\omega,T)]$ producing dispersive frequency shifts and $Im[\varepsilon_{TLS}(\omega,T)]$ contributing to dielectric loss ($1/Q_i$), while thermally or non-thermally generated quasiparticles reduce the Cooper pair density, increasing kinetic inductance ($L_k$) and lowering the resonant frequency ($f_r \propto 1/\sqrt{L_k}$). Power-independent losses from residual quasiparticles, surface defects, and trapped magnetic flux further contribute to baseline dissipation.[22,29,76] To systematically probe these effects, we measured the resonant frequency ($f_r$) and $Q_i$ of the resonators across a temperature range from 100 mK to 900 mK at three distinct input powers (-150, -120 and -90 dBm). These measurements allow us to track the evolution of both $f_r$ and $Q_i$ with temperature and excitation power, thereby providing deeper insight into the relative dominance and interplay of different loss channels under varying experimental conditions.

2.2.2 *Temperature dependent $\Delta f_r / f_r$ measurements*

At a low input power of -150 dBm, at low temperatures, TLS defects already alter the dielectric response in a temperature-dependent manner, producing a characteristic fractional shift in the resonator's frequency. However, as the temperature approaches and exceeds ~500 mK, a pronounced increase in the relative frequency change emerges (**Figure 5**). This behavior is consistent with the expected thermally activated generation of quasiparticles which produces a substantial red shift in the resonance frequency across all resonator devices. The fractional resonant frequency shift ($\Delta f_r / f_r$) can be expressed as sum of contributions from TLS and quasiparticle-induced changes (equation 2):[34,75,76]

$$\left(\frac{\Delta f_r(T)}{f_r}\right) = \left(\frac{\Delta f_r(T)}{f_r}\right)_{TLS} + \left(\frac{\Delta f_r(T)}{f_r}\right)_{QP} \qquad \ldots 2$$

where, $\left(\frac{\Delta f_r(T)}{f_r}\right)_{TLS}$ is the contribution from the TLS channel and $\left(\frac{\Delta f_r(T)}{f_r}\right)_{QP}$ is the contribution from the QP channel, both defined as follows, respectively:

$$\left(\frac{\Delta f_r}{f_r}\right)_{TLS} = \frac{1}{\pi Q_{TLS}^0}\left[Re\,\Psi\left(\frac{1}{2}+\frac{i\hbar\omega}{2\pi k_B T}\right) - \ln\left(\frac{\hbar\omega}{2\pi k_B T}\right)\right] \qquad \ldots 2a$$

$$\left(\frac{\Delta f_r}{f_r}\right)_{QP} = -\frac{\alpha}{2}\frac{\Delta L_k}{L_k} = -\frac{\alpha}{2}\frac{\Delta}{k_B T}\left[\sinh\frac{\Delta}{k_B T}\right]^{-1} \qquad \ldots 2b$$

Herein, $\Psi$ is the digamma function, $L_k$ is the kinetic inductance, $\Delta$ is the superconducting gap at zero temperature and $\alpha$ is the fraction of the kinetic inductance with respect to the total inductance.

To identify the dominant loss channels at different temperatures for the $\Delta f_r/f_r$ shift, we here introduce normalized weights ($w_{TLS}$ and $w_{QP}$), assigned to each term on the right hand-side of equation 2, after division (normalization) of the equation by $\left(\frac{\Delta f_r(T)}{f_r}\right)$. The weights are therefore constrained such that $w_{TLS} + w_{QP}=1$. Since equation 2 has only two independent parameters ($Q_{TLS}^0$ and $\alpha$), this formulation facilitates a robust extraction of $Q_{TLS}^0$. Figure 5 presents the fitting of $\Delta f_r/f_r$ against the temperature using equation 2, along with the normalized weights for the different contributions.

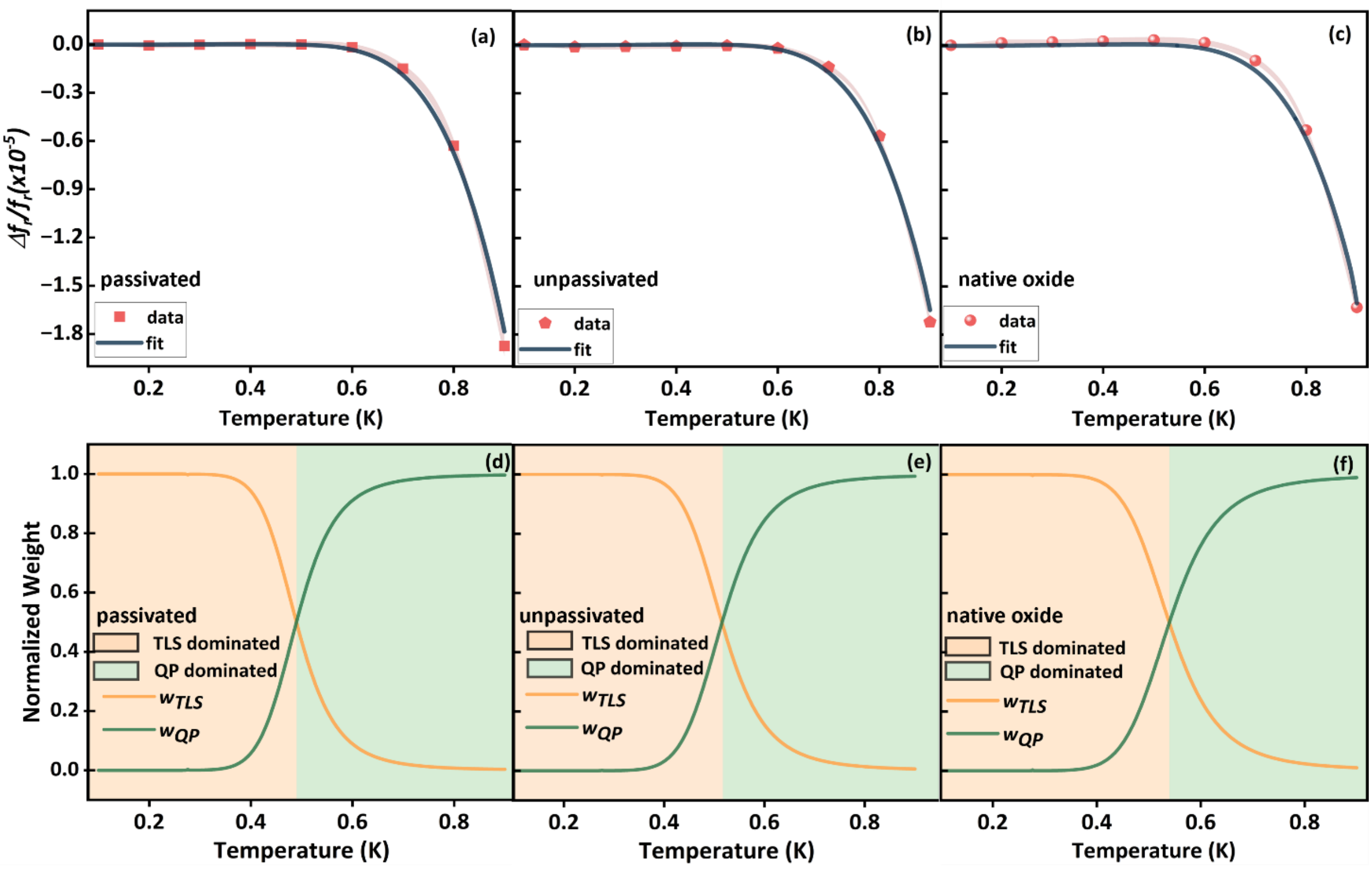

***Figure 5.*** *Mean fractional frequency shift ($\Delta f_r/f_r$) as a function of temperature, measured at excitation powers of -150 dBm of (a) passivated Ta resonators (b) unpassivated resonators and (c) native oxide resonators; the solid line represents the fit to the averaged data (red dots) using eq. 2; the thin light-red shaded region shows the 95% confidence interval of the data for different resonators. The corresponding weights characterizing the different loss contributions for passivated resonators, unpassivated resonators and native oxide resonators are displayed in d), e) and f), respectively. All normalized weights show a clear dominance of TLS contributions for lower temperatures and QP-dominated frequency shifts at higher temperatures. Note that from (d) through (e) to (f), the transition (cross-over) between both regions of dominance shifts towards higher temperatures.*

Across all devices, the resonant frequency exhibits a characteristic behavior with temperature, consistent with growing thermal saturation of TLSs upon increasing the temperature from 100 mK to several hundred mK (visible as a weak increase), followed by a strong, monotonic decrease above ~500 mK, as quasiparticle generation increases the kinetic inductance. With increase in temperature, the rapid growth in quasiparticle density produces this pronounced frequency lowering. Strikingly, native oxide resonators show a substantially larger initial frequency change than either unpassivated or passivated devices, indicating that their temperature response is dominated by TLS-mediated dielectric contributions, in agreement with the weighted loss-channel analysis as discussed further below. The evolution of the weights highlights the crossover between the dominant loss channels: from dominating TLS contribution to QP contribution with increase in temperature, due to an increased thermal QP density in the resonators. We also note that passivated resonators undergo a transition from TLS-dominated to QP-dominated behavior at lower temperatures compared to unpassivated and even more, native oxide resonators. This shift can be attributed to the reduced TLS losses in passivated devices, which in effect let quasiparticle losses dominate at already lower temperatures. This trend becomes directly evident in the comparison presented in **Figure 6**.
Finally, **Figure 7** summarizes all $Q^0_{TLS}$ and $\alpha$ values as extracted from the fits to equation 2, for all different resonator types. We obtain the highest $Q^0_{TLS}$ value for the passivated resonators, followed by the unpassivated and native oxide resonators, as could be already anticipated from the low-power behavior presented in Figure 4. In contrast, the actual surface treatment has apparently low, or no significant, impact on the kinetic inductance fraction $\alpha$: all resonators show a similar value of about 0.016-0.018 (see, Figure 7b).

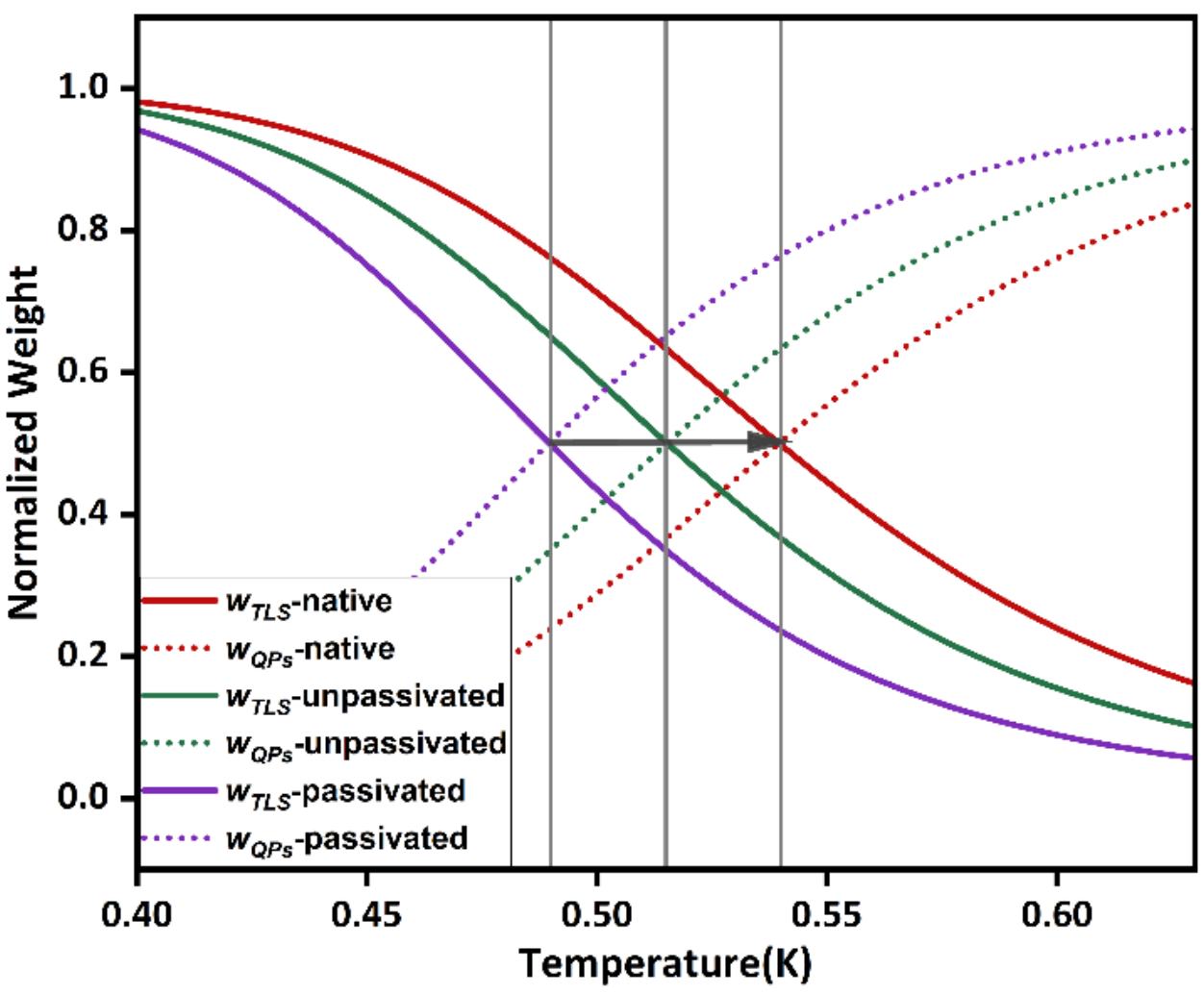

***Figure 6**. Normalized weights for $\Delta f_r / f_r$ against temperature at -150 dBm for different resonators; a clear difference in transition from a TLS dominated to a QP dominated region is observable. Passivated resonators undergo a transition from TLS-dominated to QP-dominated behavior at lower temperatures compared to unpassivated and native oxide resonators, the black arrow denotes the shift in transition between TLS and QP dominated losses for different resonators.*

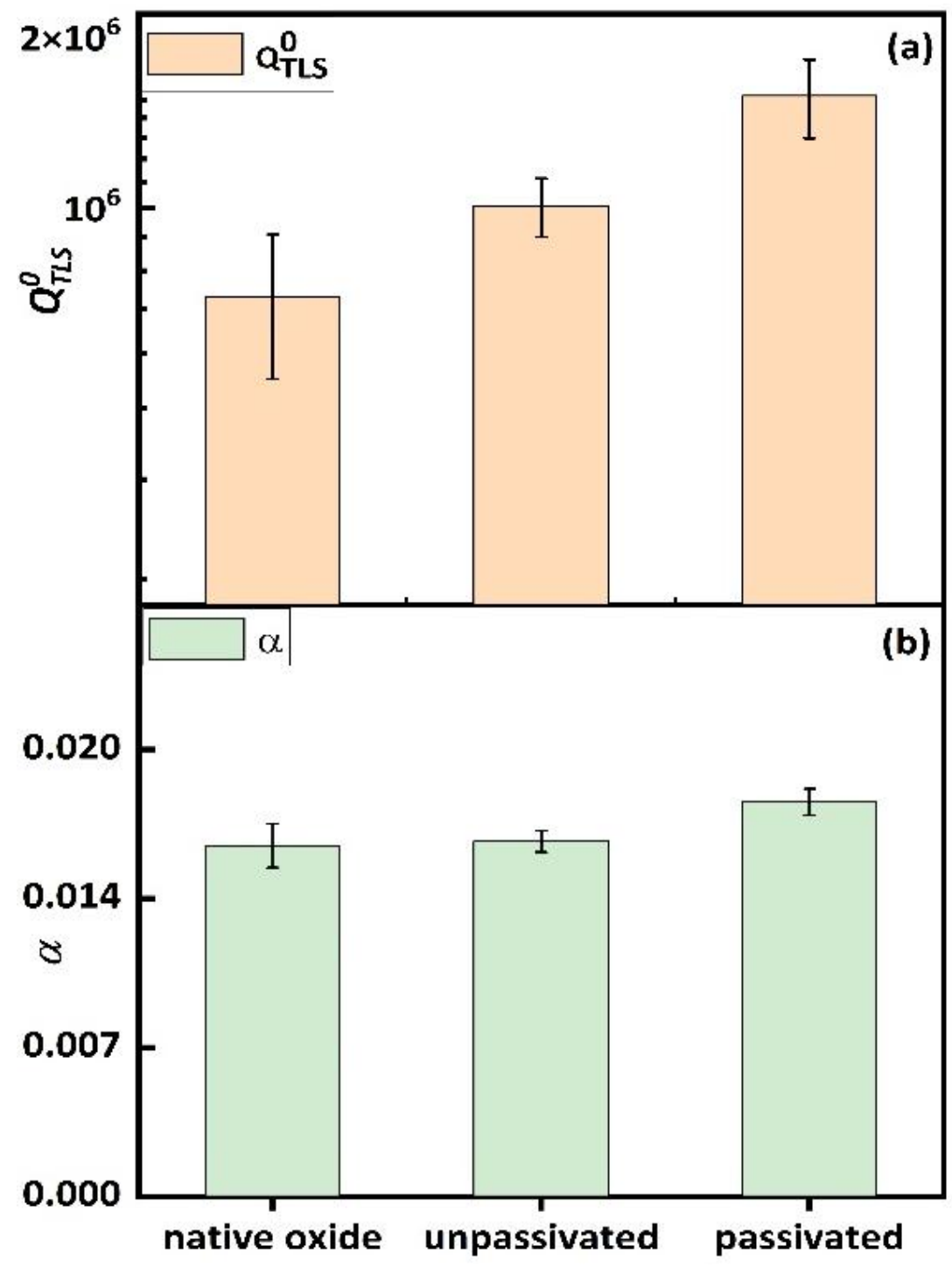

***Figure** 7. Parameters $Q^0_{TLS}$ and α, extracted for all resonator devices from the fitting of fractional frequency shifts against temperature, according to eq. 2: (a) $Q^0_{TLS}$, passivated resonators show highest $Q^0_{TLS}$ values indicating least losses due to TLS. (b) α, all resonators show a similar kinetic inductance fraction Error bars denote the standard deviation obtained by fitting from averaging over all resonators.*

*2.2.3 Temperature dependent $Q_i$ measurements*

**Figure 8** shows the variation of $Q_i$ with temperature at different excitation powers for all three resonator types. At low excitation powers (-150 dBm), all resonators exhibit an initial increase in $Q_i$ with rising temperature, consistent with the growing thermal saturation of TLSs, which dominates the losses in the low-temperature range. At higher temperatures, quasiparticle excitations in the superconducting film become the primary loss mechanism, resulting in a pronounced decrease of $Q_i$ as the system approaches the superconducting transition. To further differentiate the different loss mechanisms, we describe the resonator losses over a wide range of temperatures and microwave powers with the following model. The overall $Q_i$ for resonators can be expressed as per equation 3:

$$\frac{1}{Q_i(n,T)} = \frac{1}{Q_{TLS}(n,T)} + \frac{1}{Q_{QP}(T)} + \frac{1}{Q_{other}} \quad \ldots 3$$

where, $Q_{TLS}(n,T)$, $Q_{QP}(T)$, and $Q_{other}$ are the contributing quality factor parameters associated with TLS loss, quasiparticle loss and other, power-independent loss. $Q_{TLS}(n,T)$ and $Q_{QP}(T)$ can be defined as follows:[75]

$$\frac{1}{Q_{TLS}(n,T)} = \frac{1}{Q^0_{TLS}}\left[\frac{\tanh\left(\hbar\omega/2k_BT\right)}{\sqrt{1+\left(n^{\beta}/D_T\eta\right)\tanh\left(\hbar\omega/2k_BT\right)}}\right] \quad \ldots 3a$$

$$\frac{1}{Q_{QP}(T)} = \frac{1}{Q_{QP}^0}\left[\frac{\sinh\left(\hbar\omega/2k_BT\right) K_0\left(\hbar\omega/2k_BT\right)}{e^{\left(\Delta_0/k_BT\right)}}\right] \qquad \ldots 3b$$

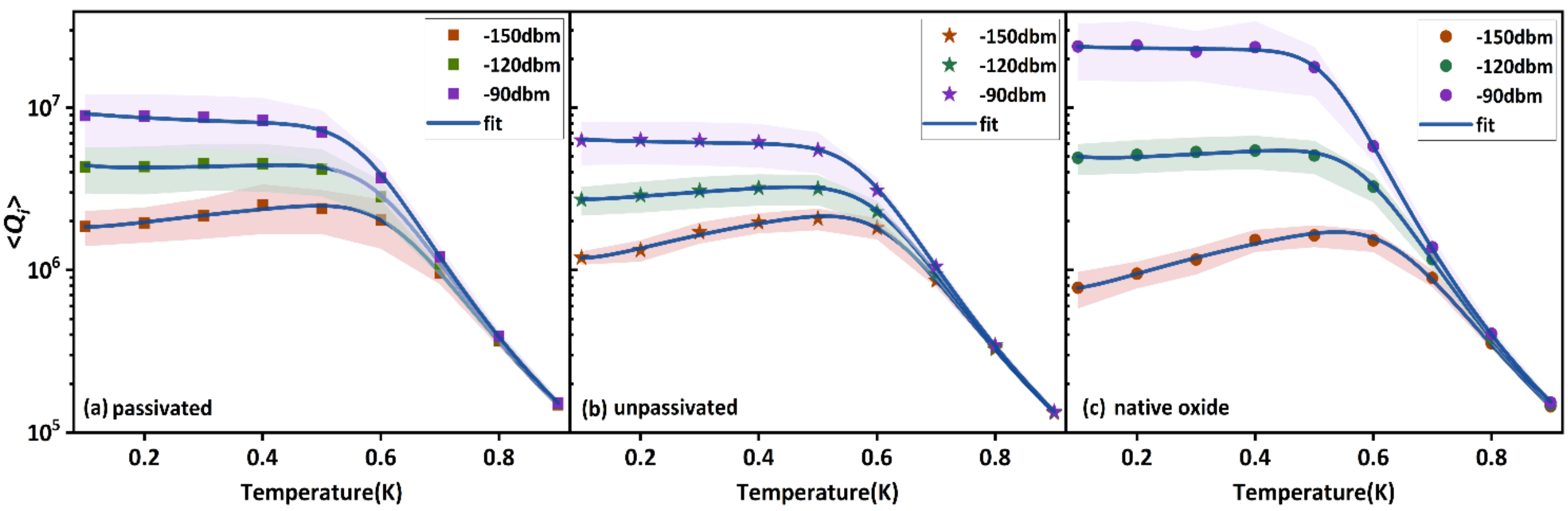

***Figure 8**. Mean internal quality factor ($Q_i$) of Ta resonators as a function of temperature, measured at excitation powers of -150 dBm, -120 dBm, and -90 dBm for (a) passivated resonators (b) unpassivated resonators and (c) native oxide resonators; the solid lines represent the fits to the data using equation 3, and the shaded regions show the 95% C.I. of the data for different resonators.*

Herein, $D$, $\eta$ parametrize the TLS saturation with temperature and $\beta$ parametrizes the saturation with input power, $Q_{QP}^0$ is the overall characteristic quasiparticle loss and $K_0$ is the zeroth order modified Bessel function. Consistent with the trend observed in the $\Delta f_r/f_r$ fits, we find that passivated resonators exhibit the highest $Q_{TLS}^0$, followed by unpassivated and native-oxide resonators (**figure 9a**), indicating reduced TLS-related losses in the passivated devices. Notably, although the ordering of $Q_{TLS}^0$ extracted from both $\Delta f_r/f_r$ $(T)$ and $Q_i$ $(T)$ analyses is the same, the values obtained from the $Q_i$ $(T)$ analysis are typically 10-20% lower. This discrepancy likely arises because the $\Delta f_r/f_r$ $(T)$ method is also sensitive to off-resonant TLSs that predominantly contribute to frequency shifts rather than to internal loss, leading to a systematically higher inferred $Q_{TLS}^0$.[76] On the other hand, the parameter $Q_{other}$, inverse of which accounts for power-independent losses, shows its highest value for native oxide resonators (figure 9b), consistent with the observation made for the $Q_i(n)$ measurements discussed previously. In superconducting resonators, $Q_{other}$ reflects the inverse of a residual, power-independent loss floor that persists once TLS losses are saturated at high excitation powers. The observed degradation of $Q_{other}$ following BOE exposure for both passivated and unpassivated resonators therefore points to an increase in non-TLS loss channels or power-independent losses. The lack of improvement with surface passivation along with no statistically significant difference in $Q_{other}$ between passivated and unpassivated resonators further supports the interpretation, that these losses do not arise from surface TLSs associated with native or regrown oxides. Instead, these are consistent with the introduction of bulk or interface-related non-TLS losses presumably induced during BOE etching. They may have originated from hydrogen related defects within the tantalum films during this step, which accompanies native oxide removal.[62,74] In this work, we use 'hydrogen related defects' as a shorthand for generic hydrogen associated defects that can form in sputtered Ta films, namely surface Ta-H terminations after oxide removal, interstitial H within the Ta lattice and hydrogen-rich sub-stoichiometric TaOx. Such configurations may contribute with non-TLS loss channels (e.g., non-saturable dielectric or resistive losses), which would differ from canonical TLSs in both their saturation behavior and temperature dependence. We thus attribute the excess loss observed in the high-power regime primarily to these non-TLS mechanisms that elevate the residual loss floor, hence reducing $Q_{other}$.

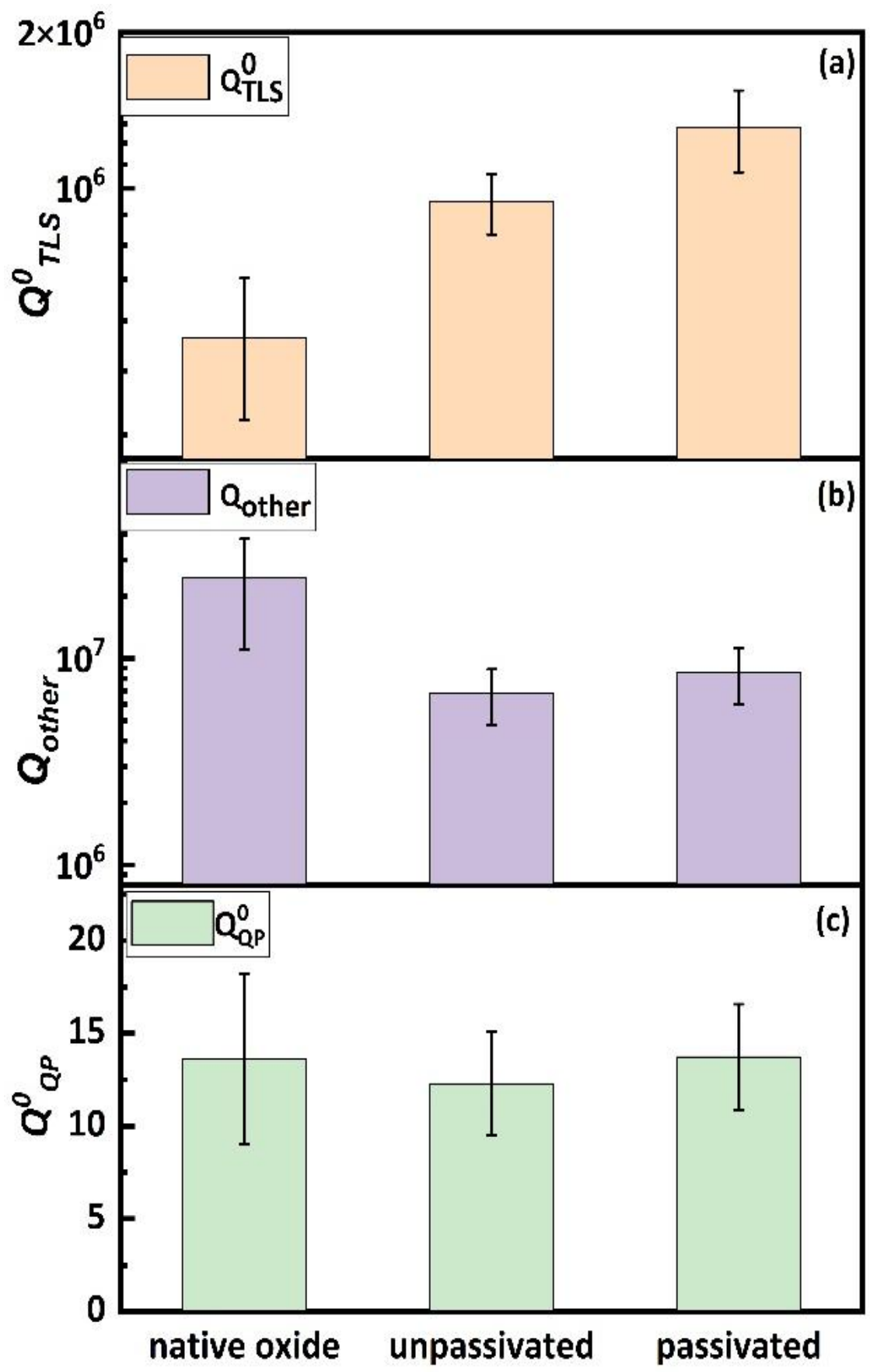

***Figure 9****. Loss-related quality factor parameters extracted for all resonator devices from the fitting of $Q_i$ against temperature at different powers, according to eq. 3: (a) $Q^0_{TLS}$, passivated resonators show highest $Q^0_{TLS}$ values - a trend similar to the one obtained using fractional frequency shift fitting, see Fig. 7. (b) $Q_{other}$, native oxide resonators show highest values, suggesting least power independent losses. (c) $Q^0_{QP}$, all resonators show similar characteristic quasiparticle loss. Error bars denote the standard deviation obtained by fitting from averaging over all resonators.*

However, we do not observe any significant difference in the parameter $Q^0_{QP}$, which accounts for the primary magnitude of quasiparticle losses, indicating no notable change in superconducting properties post surface passivation.

As discussed for the frequency shifts above, also evaluating the $Q_i$ data with normalized weights can give deeper insight into the evolution of different contributing loss mechanisms at different temperatures and powers. Hence, we equally assign normalized weights ($w_{TLS}$, $w_{QP}$ and $w_{other}$) to each term on the right hand-side of equation 3, such that $w_{TLS} + w_{QP} + w_{other} = 1$, i.e., a higher value of $w$ indicates a higher relative contribution. We find that all resonators are dominated by TLS losses at low power and low temperature, while QP losses become increasingly significant as soon as the TLSs saturate and thermally generated QPs emerge at higher temperatures, as shown in **Figure 10**.

Although the passivated resonators exhibit the highest $Q_{TLS}^{0}$, the extracted TLS weights indicate that device performance at low power and low temperature remains limited primarily by TLS dissipation. As the drive power increases, the saturation of TLSs progressively reduces their contribution, allowing other loss channels to dominate; at -90 dBm, these non-TLS mechanisms, mainly $Q_{other}$ (power-independent losses) become the leading source of dissipation at low temperatures. We also find that the onset of QP-dominated loss shifts to lower temperatures as the drive power increases (Figure S9). This trend is consistent with microwave-induced generation of additional quasiparticles, whose density is expected to scale approximately with the square root of the input power.[77] A more detailed investigation would require measurements at higher powers and with finer power resolution, which however lies beyond the scope of the present work.

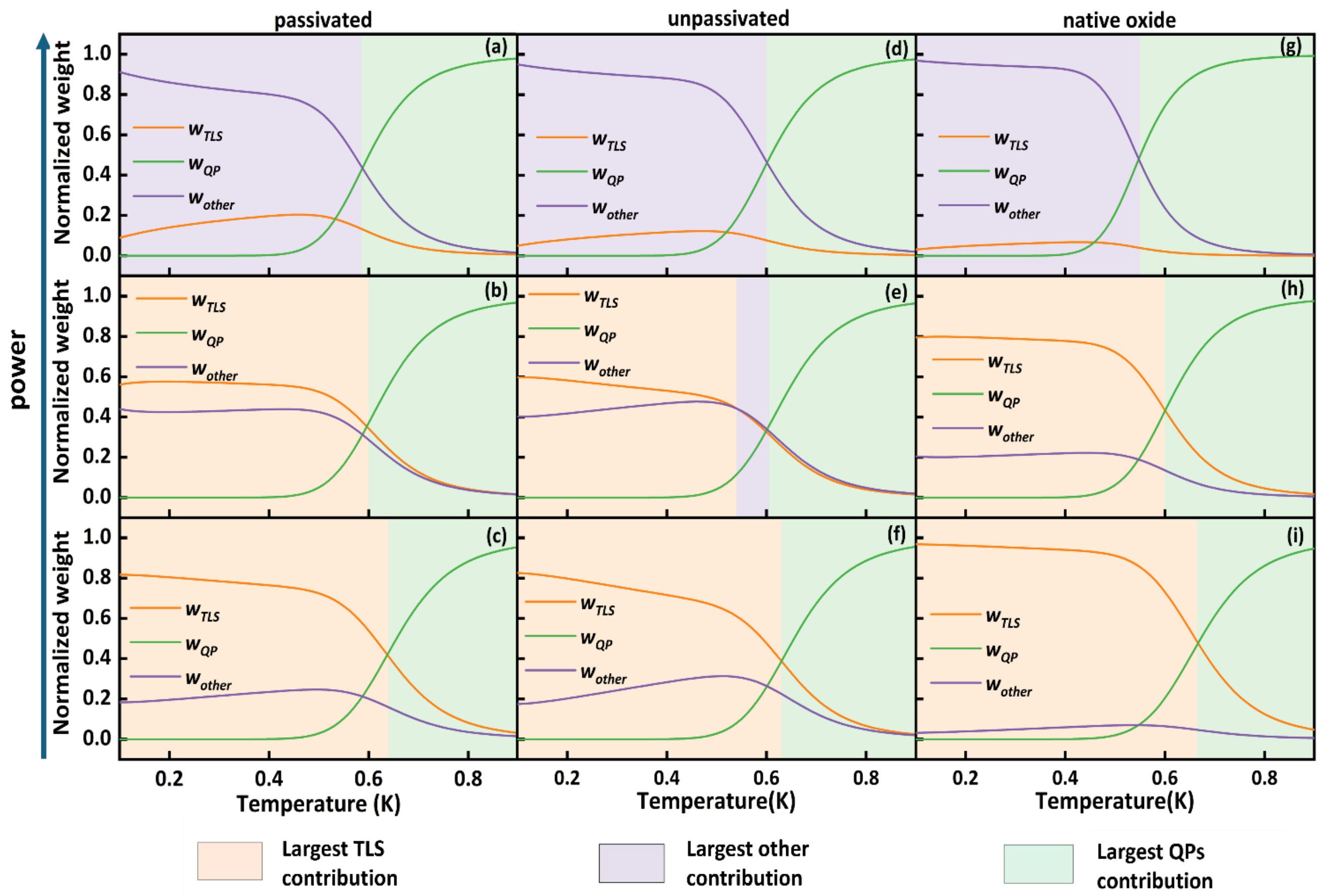

***Figure 10**. (a), (b) and (c) show the normalized weights of $Q_i$ as a function of temperature, for passivated resonators at -90 dBm, -120 dBm and -150 dBm, respectively. (d), (e) and (f) show the normalized weights of $Q_i$ for unpassivated resonators at -90 dBm, -120 dBm and -150 dBm and (g), (h) and (i) show the normalized weights of $Q_i$ for native oxide resonators at -90 dBm, -120 dBm and -150 dBm, respectively. For all resonators, the normalized weights show a clearly largest contribution of the TLS at lower temperatures and low powers, and a largest*

*QP contribution at higher temperatures, whereas with increasing power other contributions start to increase at low temperatures, as TLSs starts to saturate.*

## 3. Conclusion

In summary, we have demonstrated the simultaneous passivation of Si-air and Ta-air interfaces on superconducting resonator chips using organic self-assembled monolayers. This approach yields a marked enhancement in device performance, with passivated Ta resonators exhibiting substantially higher quality factors in the single-photon regime compared to their unpassivated or native-oxide counterparts. The improvement is driven by a reduction in TLS-induced losses, enabled by the increased hydrophobicity of the treated surfaces and the ability of the monolayer to suppress oxide regrowth, thereby limiting the re-accumulation of additional TLS defects. Importantly and in contrast to alternative surface coating or capping techniques such as PVD, our conformal deposition of self-assembled monolayers is anticipated to fully passivate all exposed resonator chip surfaces, including the side edges of the tantalum CPW structures. Future work shall be focused on a further improvement of the passivation methodology and may include additional surface treatments post monolayer growth (such as annealing), as our data suggests additional power-independent losses, potentially introduced by the organic coating process and thereby possibly related to hydrogen defects. Beyond establishing a viable route to mitigate TLS losses, our results highlight the potential of molecular interface engineering as a scalable strategy for advancing superconducting quantum devices and improving coherence in next-generation quantum technologies.

## 4. Experimental Section/Methods

### *Tantalum thin-film deposition and CPW resonator fabrication*

Tantalum thin films were deposited on high-ohmic (>10k Ohm.cm) silicon (100) substrates. Prior to deposition, the silicon chips were cleaned by ultrasonication in acetone and isopropanol for three minutes, each. After that, the substrates were dipped for 60 s into buffered oxide etch (BOE, 7:1, HF: $NH_4F$) ***[Caution:** Hydrofluoric acid is extremely hazardous and must be handled with appropriate protective equipment and training]* to remove the native oxide on the silicon, followed by rinsing with DI water and drying with $N_2$. After the chemical cleaning steps, the substrates were mounted into the chamber of an RF-magnetron sputtering tool (Alcatel, model A 450) with base pressure below $3x10^{-7}$ mbar. First, a seed layer of Nb ~5 nm was deposited on the substrate to favor the growth of α-phase tantalum. Tantalum (200 nm) was then deposited with 20 sccm Ar-flow, 18 µbar process pressure and a power of 200 W. To carry out RF characterization, quarter-wave coplanar waveguide (CPW) resonators were fabricated from the deposited Ta thin films (for details, see Supporting Information: thin film growth and fabrication).

### *Microwave measurements*

We investigated the RF response of Ta resonators by performing frequency domain measurements at varying input powers (-90 to -160 dBm, with the power in dBm obtained from the power $P_{\mathrm{mW}}$ in mW using $P_{dBm} = 10log_{10}(P_{\mathrm{mW}}/1\ \mathrm{mW})$) and varying temperatures (100-900 mK). The samples were mounted into an adiabatic demagnetization cryostat (kiutra, L-Type Rapid). The measurements involved recording the transmission coefficient $S_{21}$ for the resonators using a vector network analyzer (for details, see Supporting Information: Cryogenic resonator measurement setup, and Microwave measurement procedure and analysis).

***DC electrical measurements***

For DC electrical characterization (critical temperature, $T_c$, and residual resistance ratio, RRR), the Ta thin films were lithographically defined into a standard Hall-bar geometry. Four-terminal resistance measurements were performed using a Keithley 2420 source meter. For this purpose, the device was wire-bonded to a low-temperature sample holder, and measurements were carried out under continuous temperature control provided in a kiutra L-Type Rapid cryostat. During each measurement, a constant DC excitation current was sourced through the longitudinal Hall bar contacts while the resulting voltage drop was recorded at contacts along the bar's side, thereby eliminating contributions from lead and contact resistances.

***SAM growth and characterization***

Growth

We have grown SAMs onto three different films: (i) planar Ta films and (ii) planar Si and (iii) substrates with Ta films structured as CPW resonators. 1-Octadecene ($C_{18}H_{36}$) molecules were used in all three cases (see Supporting Information: Self assembled monolayer growth).

Characterization

The grown SAMs were characterized using contact angle goniometry, ellipsometry, X-ray photoelectron spectroscopy, Fourier-transform infrared spectroscopy, AFM and transmission electron microscopy.

- Contact angle

Surface wettability was evaluated using a Dataphysics OCA15 contact angle goniometer. A 2 µL droplet of DI water was dispensed onto the sample surface via a precision dosing system, and the static contact angle (CA) was determined from images captured immediately after droplet deposition. Contact angles were analyzed using the Dataphysics software. Reported CA values represent the mean values obtained from multiple droplets measured at a minimum of four distinct locations per sample.

- Ellipsometry

Variable angle spectroscopic ellipsometry measurements were made with a J.A. Woollam Co. alpha-SE ellipsometer at two different incidence angles (75° and 65°) over a wavelength range of 400-900 nm. The thicknesses of the SAMs were estimated from ellipsometry by fitting the top layer with Cauchy's dispersion relation.

- X-ray photoelectron spectroscopy

X-ray photoelectron spectroscopy (XPS) measurements were performed using a home-built system (components from SPECS Surface Nano Analysis) operating under ultra-high vacuum (~$5\times10^{-9}$ mbar). The setup was equipped with an XR 50 X-ray source and a Phoibos 100 hemispherical electron analyzer. A magnesium anode was employed to generate Mg $K_{\alpha}$ radiation (E = 1253.6 eV). The X-ray source was operated at 12.50 kV and 20.0 mA. Photoelectron spectra were recorded in fixed analyzer transmission mode with a pass energy of 25 eV and a dwell time of 1 s. Each spectrum represents the average of two to four consecutive scans. Data analysis, including charge referencing and background subtraction, was performed using CasaXPS and OriginLab software. All spectra were charge-referenced to the C 1s hydrocarbon peak at 285.0 eV.

- Fourier-transform infrared (FTIR) spectroscopy

The structural order and quality of the SAMs was investigated by infrared spectroscopy. FTIR spectra in attenuated total reflection (ATR) mode were recorded using a Bruker spectrometer (Vertex 70V), to obtain the molecular vibration fingerprint of the SAM. FTIR spectra of the passivated samples were obtained by first measuring the background response from bare Si and Ta substrates. This background was then subtracted from the corresponding data of the passivated films to yield the final spectra.

- X-ray diffraction

To characterize the crystallographic orientation and phases of Ta, all planar films were probed with grazing angle incident-X-ray diffraction (GI-XRD), with a fixed small incidence angle of 0.5° to minimize the background signal from the bulk Si and maximize the signal from the top-layer.

- Atomic force microscopy

To determine the surface morphology and validate the conformal growth of the SAMs, AFM images were recorded in tapping mode using a Bruker (Veeco) Dimension V AFM. The surface roughness post SAM passivation of the Si and Ta thin films were measured. The images were processed and analyzed for roughness estimation using the Gwyddion 2.71 software.

- Transmission Electron Microscopy/ Electron energy loss spectroscopy

Transmission Electron Microscopy (TEM) measurements were performed on cross-sections of the thin films prepared by SGS Institut Fresenius GmbH, using a JEOL JEM F200 microscope with an operation voltage range 80-200 kV. Electron energy loss spectroscopy (EELS) spectra, to obtain elemental mapping and confirm the presence of the SAM, were obtained with the same microscope using GATAN GIF Continuum ER detectors.

**Acknowledgements**

We acknowledge funding through the Munich Quantum Valley (MQV) project by the Free State of Bavaria, Germany, and the Munich Quantum Valley Quantum Computer Demonstrators- Superconducting Qubits (MUNIQC-SC) project (grant no. 13N16188), funded by the Federal Ministry of Research, Technology and Space, Germany. The Walter Schottky Institute (Prof. Sharp, Prof. Stutzmann), the Zentrum für Nanotechnologie & -materialien and the Catalysis Research Center of TU Munich are gratefully acknowledged for providing access to their XPS, clean room and XRD facilities, respectively. We thank the staff of the ZEITlab shared facilities of TU Munich for expert technical support. Dr. Manoj Kumar from CIT-EE, TU Munich-Germany and Dr. Rui Pereira from Fraunhofer-EMFT is acknowledged for scientific discussion.

**Competing interests**

The authors declare no competing interests.

**Data Availability Statement**

The data that support the findings of this study are available from the corresponding author upon reasonable request.

# Supporting Information

**Enhanced Tantalum Superconducting Resonator Performance via All-Surface Organic Monolayer Passivation**

*Harsh Gupta[#*], Moritz Singer[#], Benedikt Schoof, Anna Cattani-Scholz, Shreya Sharma, Luca Rommeis, Marc Tornow[*]*

Harsh Gupta, Moritz Singer, Benedikt Schoof, Anna Cattani-Scholz, Luca Rommeis, Marc Tornow
*School of Computation, Information and Technology, Department of Electrical Engineering, Technical University of Munich, 85748 Garching, Germany*
E-Mail*: harsh.gupta@tum.de, tornow@tum.de

Shreya Sharma
*Department of Biosciences and Bioengineering, Indian Institute of Technology Roorkee, 247667, Roorkee, India*

Luca Rommeis
*Fraunhofer Institute for Electronic Microsystems and Solid State Technologies (EMFT), 80686 Munich, Germany*

**Thin film growth and fabrication**

All tantalum thin films were deposited on high-ohmic (>10 kOhm.cm) silicon (100) substrates from Topsil GlobalWafers A/S (Denmark). Silicon substrates were ultrasonically cleaned in acetone and isopropanol for 3 min each, followed by immersion in buffered oxide etch (BOE; 7:1, HF: $NH_4F$ = 12.5: 87.5%) for 60 s to remove the native oxide layer *[**Caution:** Hydrofluoric acid is extremely hazardous and must be handled with appropriate protective equipment and training]*. The substrates were subsequently rinsed with deionized water and dried under a nitrogen stream. The cleaned substrates were immediately loaded into an RF magnetron sputtering system (Alcatel A450) with a base pressure of $\sim 10^{-7}$ mbar for film deposition. A ~5 nm niobium seed layer was first deposited, followed by tantalum deposition, carried out at an argon flow rate of 20 sccm, a working pressure of 18 µbar and an RF power of 200 W. It should be noted that the substrates remained under vacuum after the niobium seed layer deposition to prevent surface oxidation prior to tantalum growth. Material characterization by X-ray diffraction, electrical characterization to obtain RRR and $T_c$, XPS and TEM were performed on planar films.

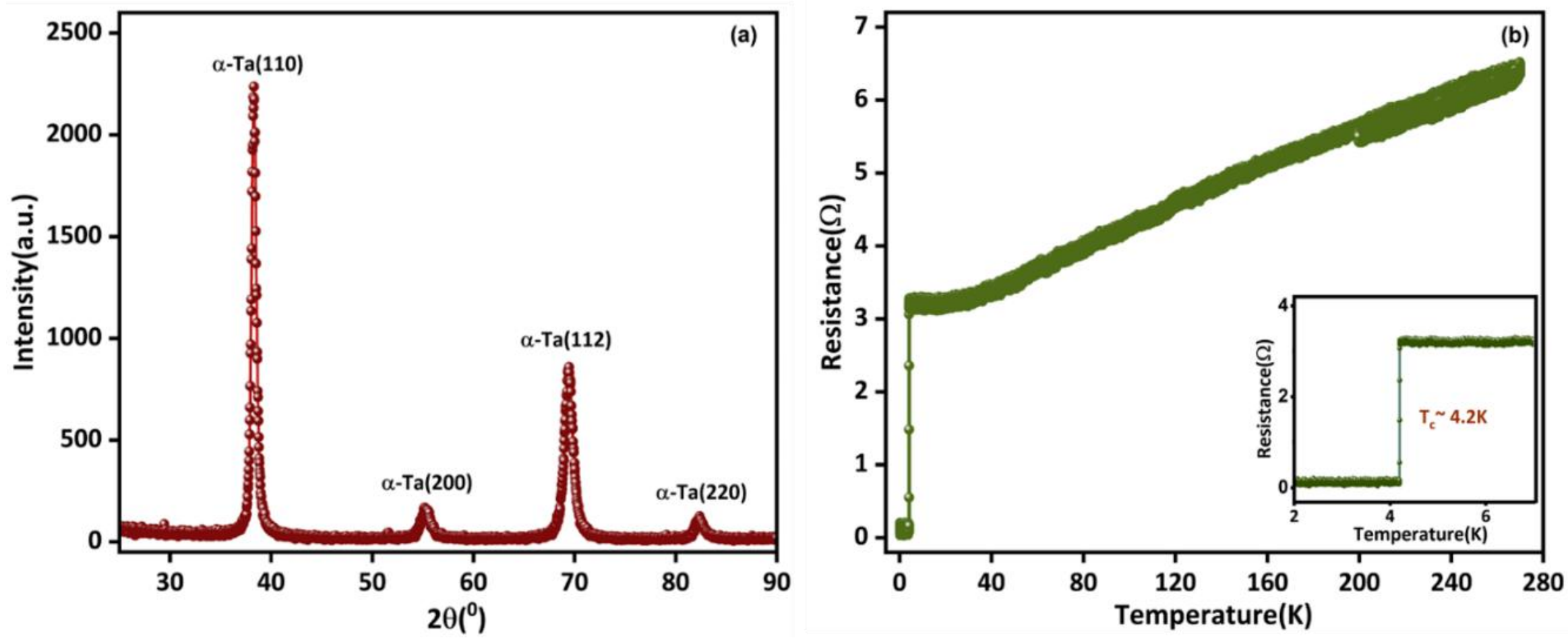

***Figure S1***. *Characteristics of α-Ta thin films grown on Si substrate with Nb seed layer (a) GI-XRD spectra for thin films; we observe only α-phase Ta peaks. (b) 4-point critical temperature measurement on an α-Ta thin film; a critical temperature of ~4.2K is recorded.*

To define the coplanar waveguide resonator structures, the desired patterns were transferred onto the chips by photolithography using a maskless aligner (μMLA, Heidelberg Instruments), followed by reactive ion etching (RIE) in an Oxford PlasmaPro 80 Cobra system, employing $CF_4$ gas (RF power: 70 W; ICP power: 150 W; etch time: 2 min). Our CPW resonators are quarter-wave resonators designed by shorting one end of a transmission line to have resonant frequencies ($f_r$) between 5-9 GHz, where $f_r$ is related to length ($l$) and effective dielectric constant ($\varepsilon_{eff}$) as [1,2]

$$f_r \propto \frac{c}{l\sqrt{\varepsilon_{eff}}}$$

where, $c$ is the speed of light in vacuum and $\frac{c}{\sqrt{\varepsilon_{eff}}} = v_{ph}$ ,is the phase velocity of light in the resonator. The typical design of the resonator chip including the dimensions of the CPW center line width and the gaps is shown in figure S2.

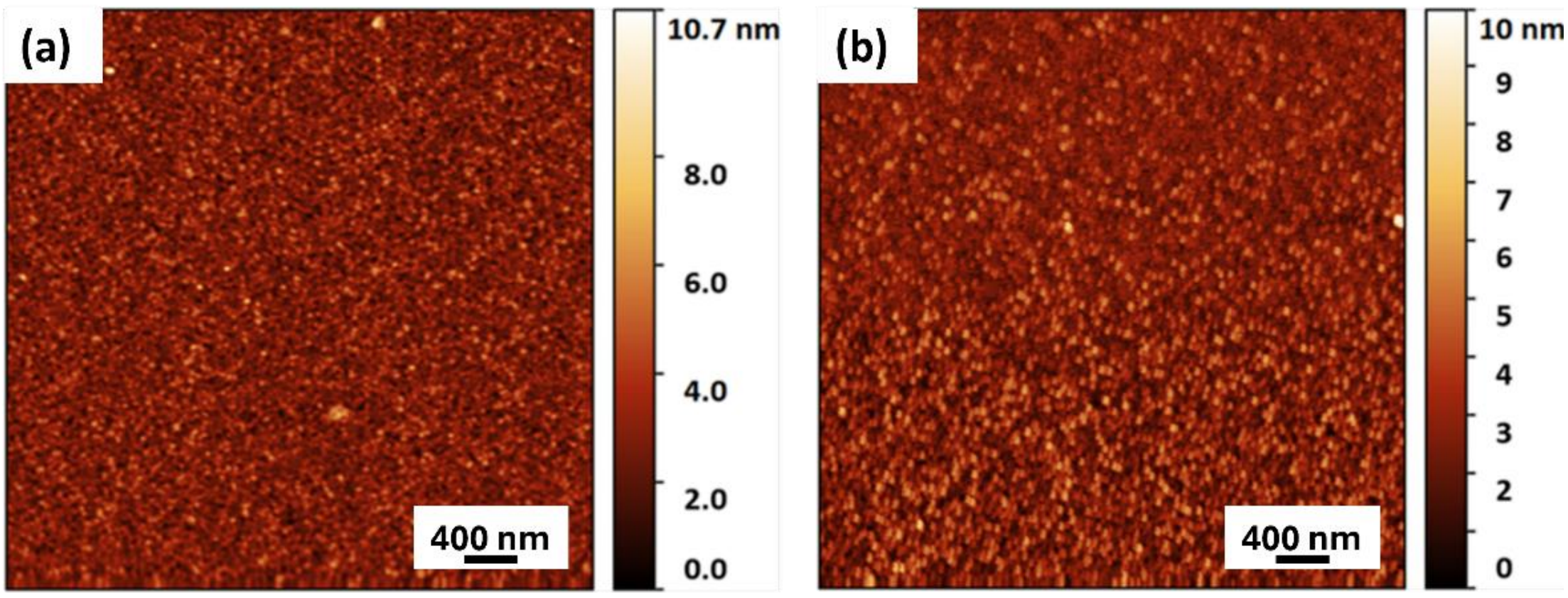

***Figure S2***. *AFM images showing the topography of (a) a SAM-passivated Si substrate with RMS roughness 0.7±0.1 nm and (b) a SAM-passivated Ta thin film with RMS roughness 0.9±0.2 nm.*

**Self-assembled monolayer growth**

The following recipe is employed for the growth of alkene SAMs on Si, as well as on Ta planar films and on Ta resonators. Thin films and resonators were first cleaned with acetone followed by isopropanol and DI water using ultrasonication for 3 min each and then etched in BOE. Subsequently, the etched samples were promptly (within ~5 mins) transferred to a reaction flask containing 1-octadecene (Sigma Aldrich), where the SAMs were then grown from the liquid phase via thermal activation, when heated for 1 hour at 160°C under nitrogen environment. After SAM growth, the samples were rinsed in fresh, clean isopropanol and dried in a nitrogen flow.

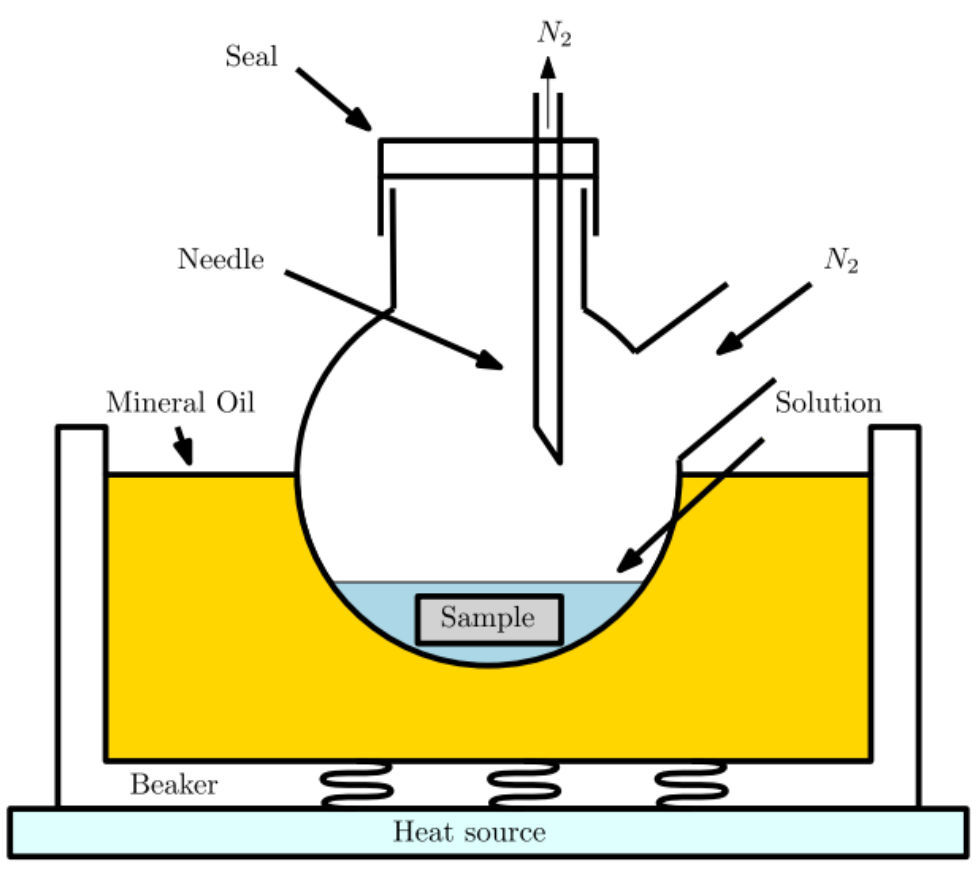

***Figure S3***. *Schematic showing the experimental setup used for the growth of alkene self-assembled monolayers on silicon and tantalum planar thin films and resonators in nitrogen ambient.*

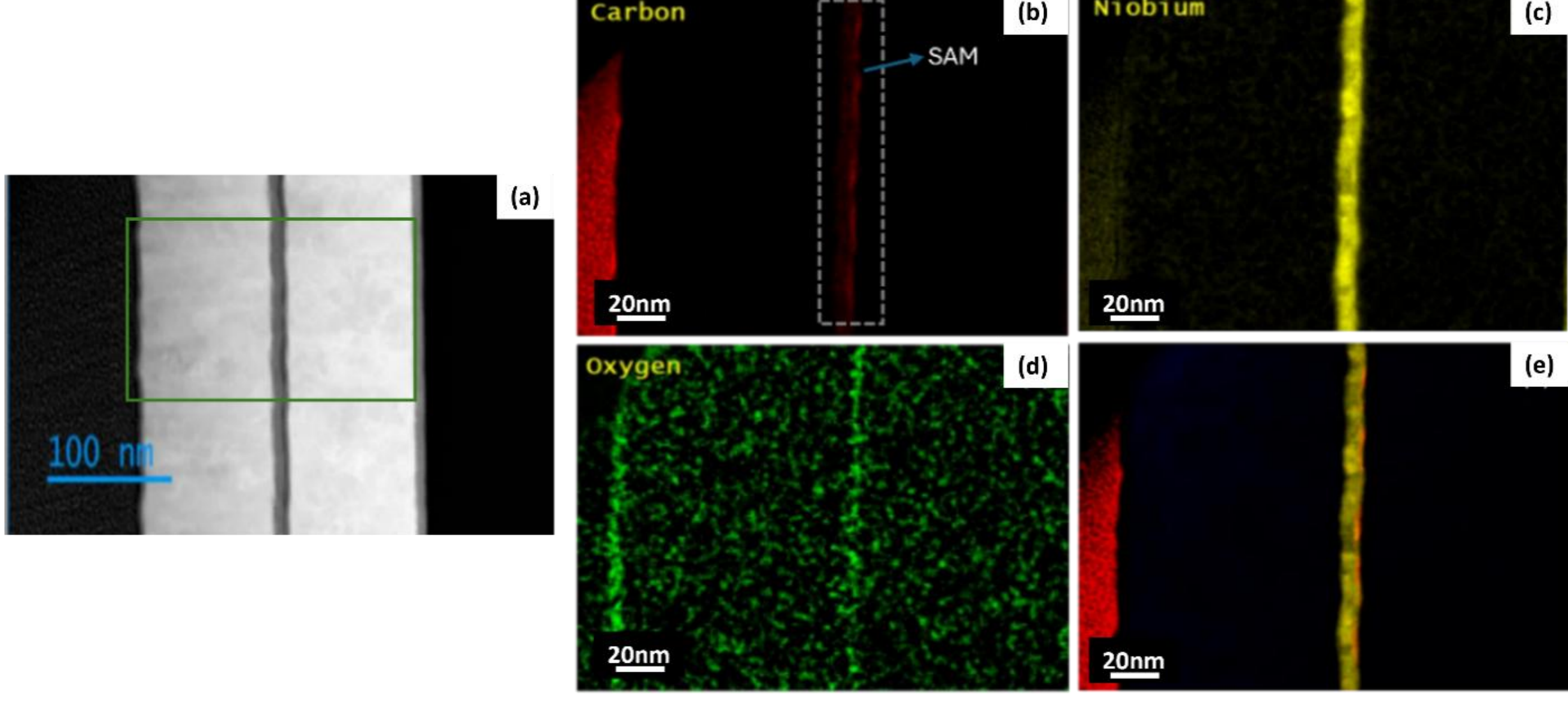

***Figure S4***. *(a) Cross-sectional TEM image and corresponding EELS maps for (b) carbon; excess carbon in center indicates the SAM, (c) niobium; the protective niobium layer on top of the SAM can be discerned. Note that this Nb layer is covered by a ca. 100 nm additional Ta layer (black region on the left of Nb), (d) oxygen; the residual Ta-oxide layer below the SAM appears prominently, and (e) cumulative EELS image.*

**Cryogenic resonator measurement setup:**
Frequency domain transmission spectra ($S_{21}$) in the microwave regime were recorded using a Keysight P5002B Streamline Vector Network Analyzer (VNA), with the sample mounted in a kiutra L-Type Rapid ADR refrigerator operating at ~100 mK. The resonator sample is housed in a copper two-port box. The RF signal was routed through wire bonds from SMA connectors located on the two-port copper housing towards the coplanar waveguide resonators. Further, to provide grounding for the resonator chip, the sample's ground plane was wire-bonded to the copper box, see Fig. 4d of the main manuscript.

On the input port, -50 dB of total attenuation across the various temperature stages of the cryostat were installed and a 20 dB attenuator is mounted on the VNA outside the cryostat. An additional damping of ~-20 dB occurred through filters, cables and connectors. Following a low pass filter (K&L microwave, cut off: 12 GHz) operating at 100 mK, the input signal is fed into the resonator chip under test. On the output line, a second low pass filter followed by a HEMT amplifier at 4 K provides the resulting $S_{21}$ response, which then goes back into the VNA and is analyzed further.

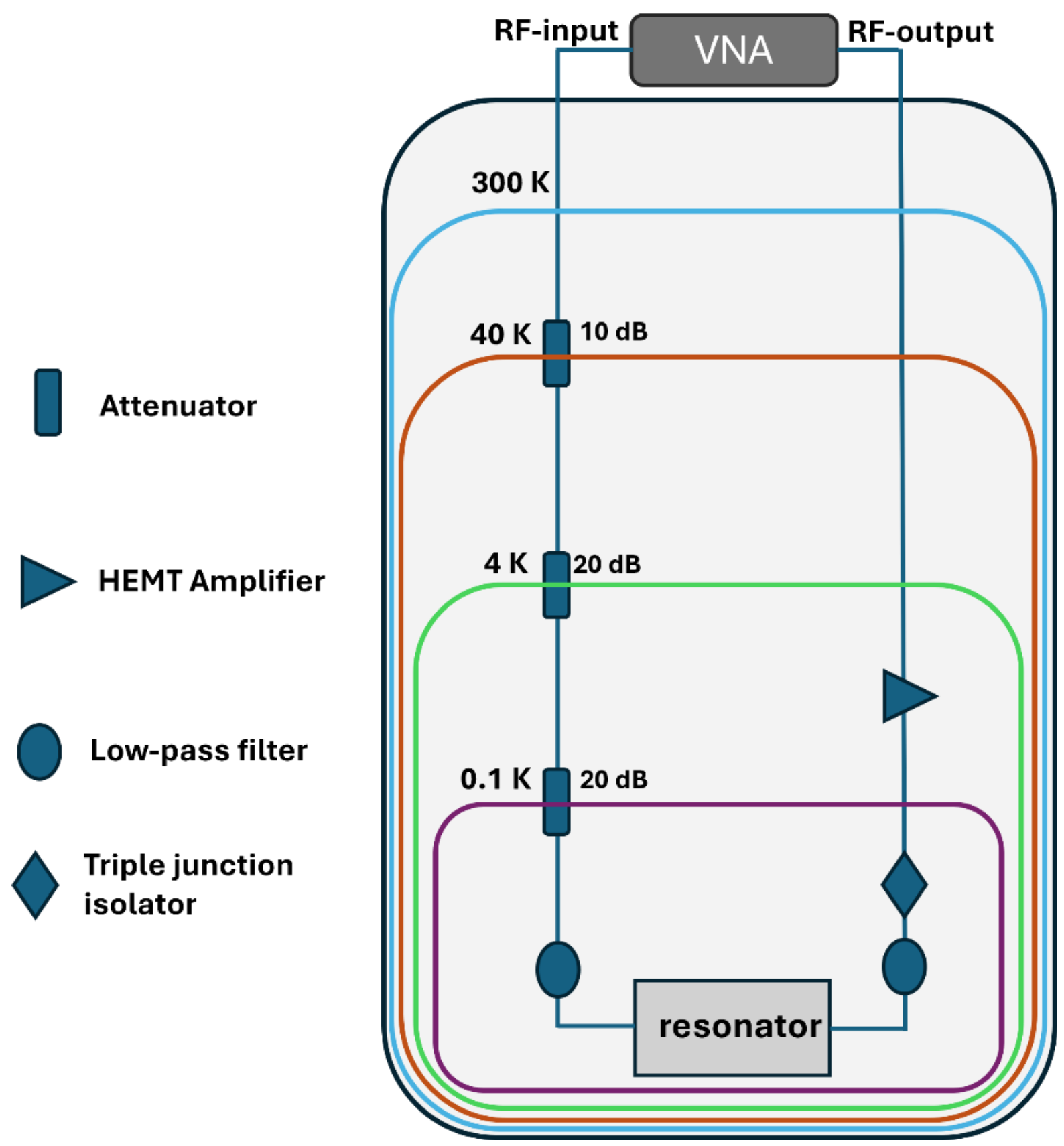

***Figure S5***. *Schematic of the experimental setup for microwave resonator measurements at cryogenic temperatures (~100 mK). The input signal from the vector network analyzer (VNA) enters the cryostat via the left RF-input port, passes through the resonator device under test, and exits through the output line. The output signal is amplified by a low-noise high electron mobility transistor (HEMT) amplifier before returning to the VNA via the right RF-output port for analysis.*

**Microwave measurement procedure and data analysis:**
We investigated the RF response of CPW resonators in transmission mode by performing frequency domain measurements at varying input powers and at different temperatures. First, a coarse frequency sweep from 4–9 GHz with 1 kHz resolution was first used to locate the resonant frequencies. Each mode was then probed with a finer 200 Hz step-size, followed by a power-dependent measurement in which the input power was varied from -90 dBm to -160 dBm in -10 dBm increments. The values for the loaded quality factors $Q_l$ and the coupling quality factors $Q_c$ were extracted, by fitting the complex-valued frequency domain $S_{21}$ data with a circle fit according to: [3,4]

$$S_{21}(f) = ae^{i\alpha}e^{-2\pi i f\tau}\left[1 - \frac{(Q_l/|Q_c|)e^{i\varphi}}{1+2iQ_l(\frac{f}{f_r}-1)}\right] \quad \text{(S1)}$$

Here, $f$ is the probe frequency, $f_r$ is the resonator frequency, and $\phi$ is the argument of the complex-valued coupling quality factor, which considers a possible impedance mismatch. To take the damping of the cables into consideration, the amplitude is modified by the factor $a$, and an initial offset of $\alpha$ is introduced in the phase along with a delay $\tau$ due to the finite length of the measurement cables. The resonator response is characterized by the $Q_l$, which combines intrinsic losses and external coupling and directly determines the resonance linewidth. Decomposing $Q_l$ into the internal quality factor $Q_i$ and coupling quality factor $Q_c$ isolates materials-limited dissipation from feedline loading. $Q_i$ reflects material related dielectric, quasiparticle and interface-related losses, whereas $Q_c$ is a geometric parameter set by the resonator feedline coupling design. Together, these parameters enable quantitative benchmarking of fabrication and surface-dependent improvements in device performance. In the complex plane, the $S_{21}$ response is a circle with diameter $\frac{Q_l}{|Q_c|}$; the loaded and coupling quality factors can be extracted by using equation S1. From them, $Q_i$ can be determined using the relation $\frac{1}{Q_i} = \frac{1}{Q_l} - \frac{1}{Q_c}$.

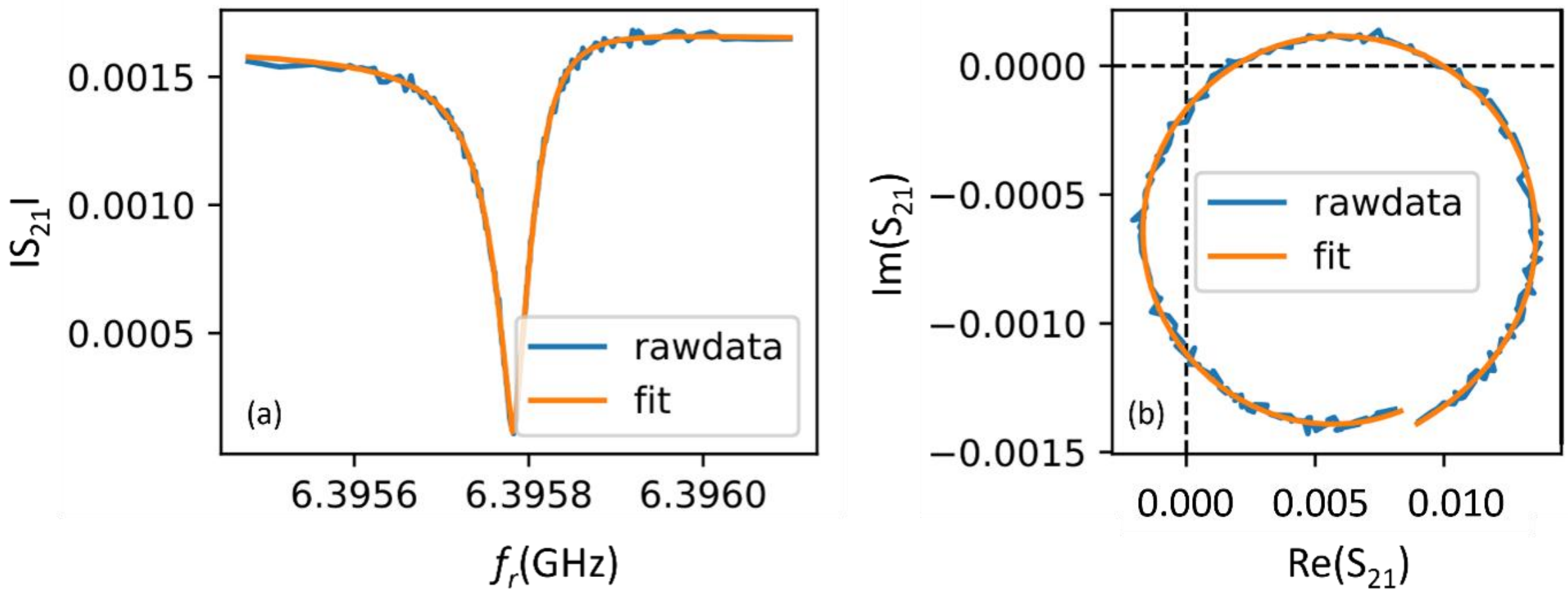

***Figure S6.*** *Exemplary resonator response traces, measured in transmission mode in the few GHz regime together with numerical fits at -150 dBm. (a) Lorentzian resonance response in the frequency domain, (b) Circular response representation in the complex $S_{21}$ plane, with a diameter $Q_l/|Q_c|$.*

The microwave power inserted into the resonator circuit can be translated into a mean photon number <*n*>, which can be estimated from the following relation, with $P_{in}$ being the input microwave power: [5,6]

$$\langle n \rangle = \frac{Q_l^2}{\pi h f_r^2 Q_c} P_{in} \tag{S2}$$

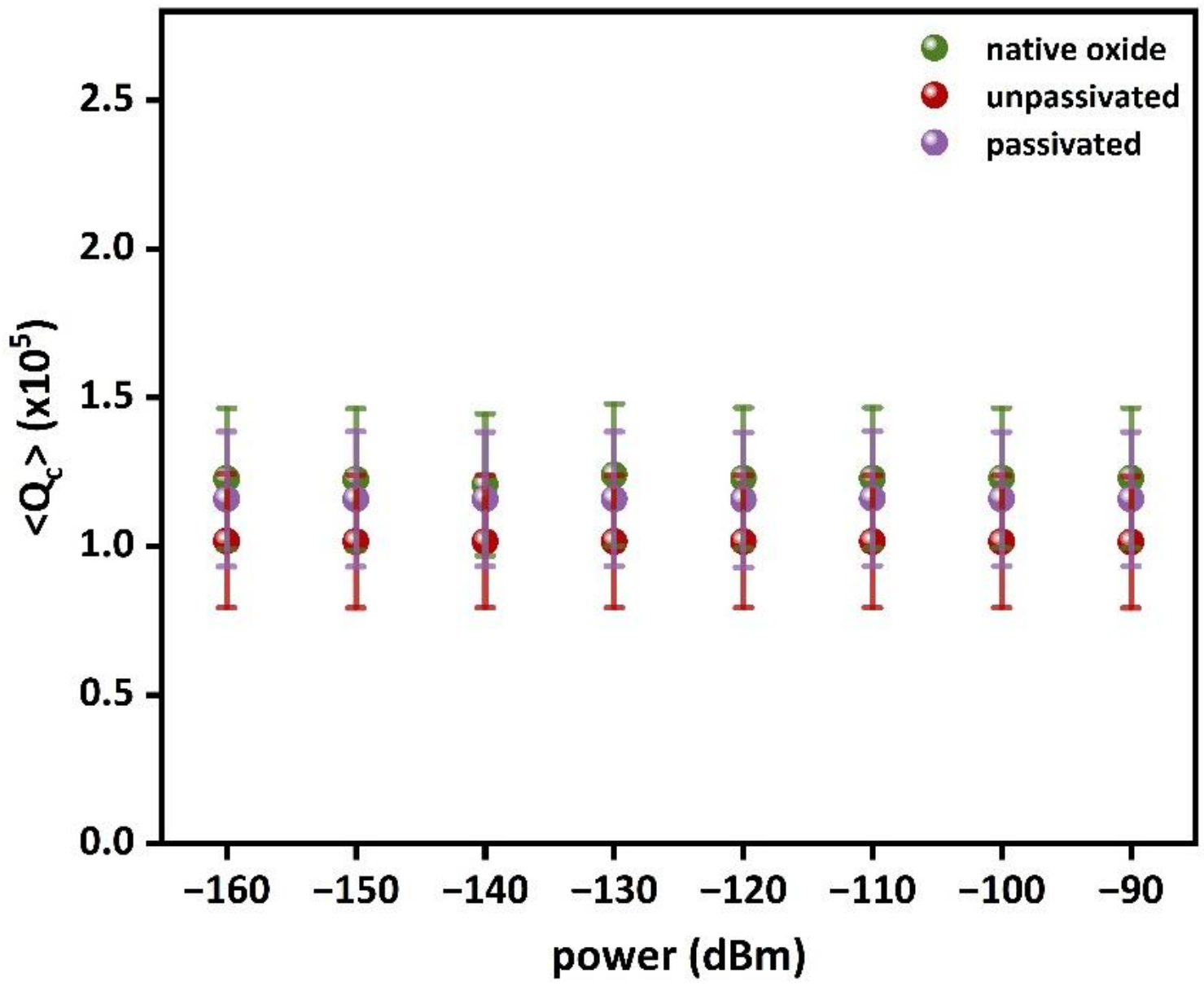

***Figure S7.*** *Coupled quality factor $Q_c$, as extracted for native oxide, unpassivated and passivated resonators. We observe no significant change in $Q_c$ with input power, also no significant change between different resonators, suggesting a robust fitting routine and confirming the similar design for all chips under investigation.*

We further validate the fitting routine; as $Q_c$ depends only on the design of the resonators, its value (stability) can confirm the extraction of parameters using equation S1. The coupling quality factor $Q_c$ is consistently extracted with a value of about ~1 - $1.5\times10^5$ (fig. S7), for resonators with native oxide, as well as for unpassivated and passivated devices. This value remains stable across varying input powers, highlighting both the reproducibility of the measurements and the reliability of the fitting procedure employed to determine $Q_i$.

Figure S8 represents the variation of extracted $Q_i$ values for all different, individual resonators (native oxide, unpassivated and passivated) at their different resonance frequencies. The passivated resonators show overall higher $Q_i$ values in the low power regime (n~1), when compared to the other resonator types, whereas native oxide resonators show overall higher $Q_i$ values in the high power regime (n~$10^6$).

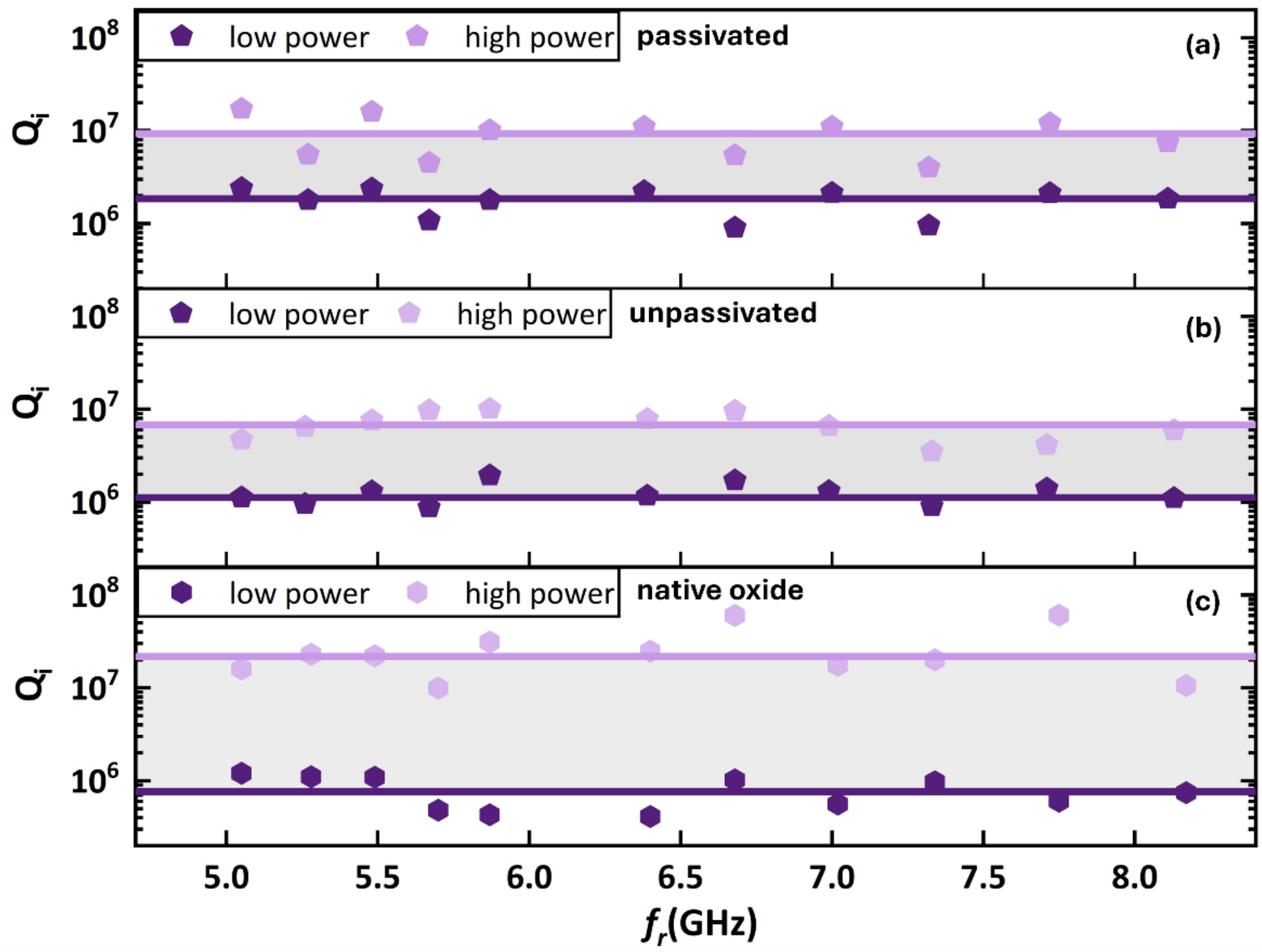

***Figure S8***. *Extracted $Q_i$ values for all single resonators with different frequencies in the range ~5-9 GHz, at both low (n~1) and high power (n~10$^6$): (a) passivated, (b) unpassivated and (c) native oxide resonators. Solid lines represent the mean values of $Q_i$ at low power (dark purple) and high power (light purple).*

**Internal quality factor fitting weights:**

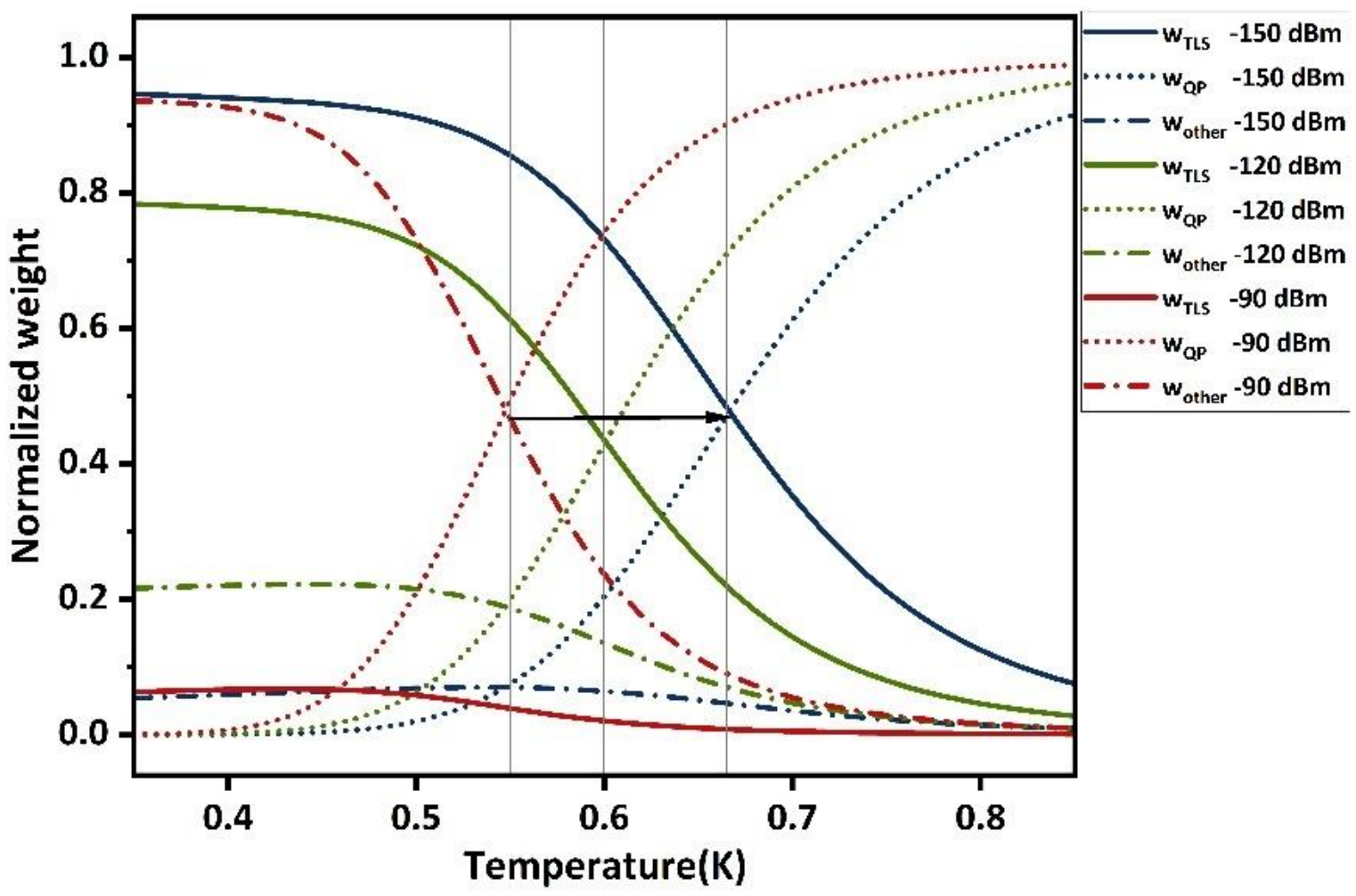

***Figure S9.*** *Normalized weights for $Q_i$ against temperature for native oxide resonators at different powers: a clear difference in transition from the TLS dominated/other dominated to the QP dominated region is observable with increase in temperature/power. The QP contribution onset shows a shift to lower temperatures with increase in power. The black arrow denotes the shift in transition between different losses for different resonators.*

**X-ray photoelectron spectroscopy:**
First, XPS survey spectra were recorded for all samples to identify all the constituent elements, in our case Ta (4f ~18-34 eV), O (1s around 532 eV), C (1s around 285 eV). Thereafter, a narrow scan spectrum for all elements was recorded. For all the spectra, a charge referencing correction was performed considering the C1s hydrocarbon peak at 285eV, no additional charge compensation/charge neutralizer was used. Fitting routine for Ta spectrum: For detailed analysis of the Ta surface chemistry, we considered the Ta-4f spectrum for all native oxide, unpassivated and passivated Ta thin films. We subtracted the Shirley background from the spectra in the binding energy range of interest. Different components were fitted with a LA ($\alpha,\beta,c$) line shape, which is a superset of the Voigt function, in CASA XPS, where $\alpha$, $\beta$ defines the spread of the tail on either side of the Lorentzian component, and the parameter $c$ specifies the width of the Gaussian used to convolute the Lorentzian curve. Different oxidation states for Ta were identified; for the fitting routine, spin-orbit doublets ($4f_{7/2,\ 5/2}$) were considered for all suboxides with fixed energy separation of ~1.9 eV and the defined area ratio of 4:3 for the $f_{7/2}$:$f_{5/2}$ peaks. The $Ta^0$ oxidation state, corresponding to metallic tantalum, was fitted using an asymmetric peak, as the presence of unfilled one-electron levels in the conduction band allows electrons to undergo shake-up type processes following core-level photoemission. Consequently, instead of distinct shake-up satellite features, a tail appears on the higher binding energy side of the main peak, resulting in an asymmetric peak shape characteristic of metallic states [7]. In contrast, such energy levels are not available in metal oxides, and therefore, the corresponding peaks exhibit symmetric line shapes. Hence, other oxidation states ($Ta^{+4}$, $Ta^{+5}$) corresponding to $TaO_2$ and $Ta_2O_5$ are fitted with symmetric peaks and with the same FWHM for individual spin-orbit doublets for different suboxides.

***Table S1*** *Peak positions, line shapes and FWHM for the different components with spin-orbit doublets of the Ta 4f XPS spectra for Ta thin films with native oxide. Similar parameters were used for passivated and unpassivated Ta thin films, within 5% deviation. Metallic Ta contributions are fitted with asymmetric peaks with parameters (α,β,c) and oxide contributions are fitted with symmetric peaks with parameters (α,β).*

| | Position(eV) | Line shape (α,β,c) | FWHM (eV) |
|---|---|---|---|
| **Ta($4f_{7/2}$)** | 21.8 | LA (1.30,2.44,69) | 1.1 |
| **Ta($4f_{5/2}$)** | 23.7 | LA (1.30,2.44,69) | 1.1 |
| **$Ta^{+5}$ ($4f_{7/2}$)** | 27.1 | LA (1.53, 2.43) | 1.5 |
| **$Ta^{+5}$ ($4f_{5/2}$)** | 29.0 | LA (1.53, 2.43) | 1.5 |
| **$Ta^{+4}$ ($4f_{7/2}$)** | 24.6 | LA (1.53, 2.43) | 2.2 |
| **$Ta^{+4}$ ($4f_{5/2}$)** | 26.5 | LA (1.53, 2.43) | 2.2 |

***Table S2*** *Peak positions, line shapes and FWHM for the different components of the C 1s XPS spectra for Ta thin films with native oxide. Similar parameters were used for passivated and unpassivated Ta thin films, within 5% deviation.*

| | Position(eV) | Line shape (a,β) | FWHM (eV) |
|---|---|---|---|
| **C-C** | 285 | LA (1.53, 2.43) | 1.2 |
| **C-OH/N** | 285.7 | LA (1.53, 2.43) | 1.9 |
| **HO-C=O** | 289.2 | LA (1.53, 2.43) | 2.3 |

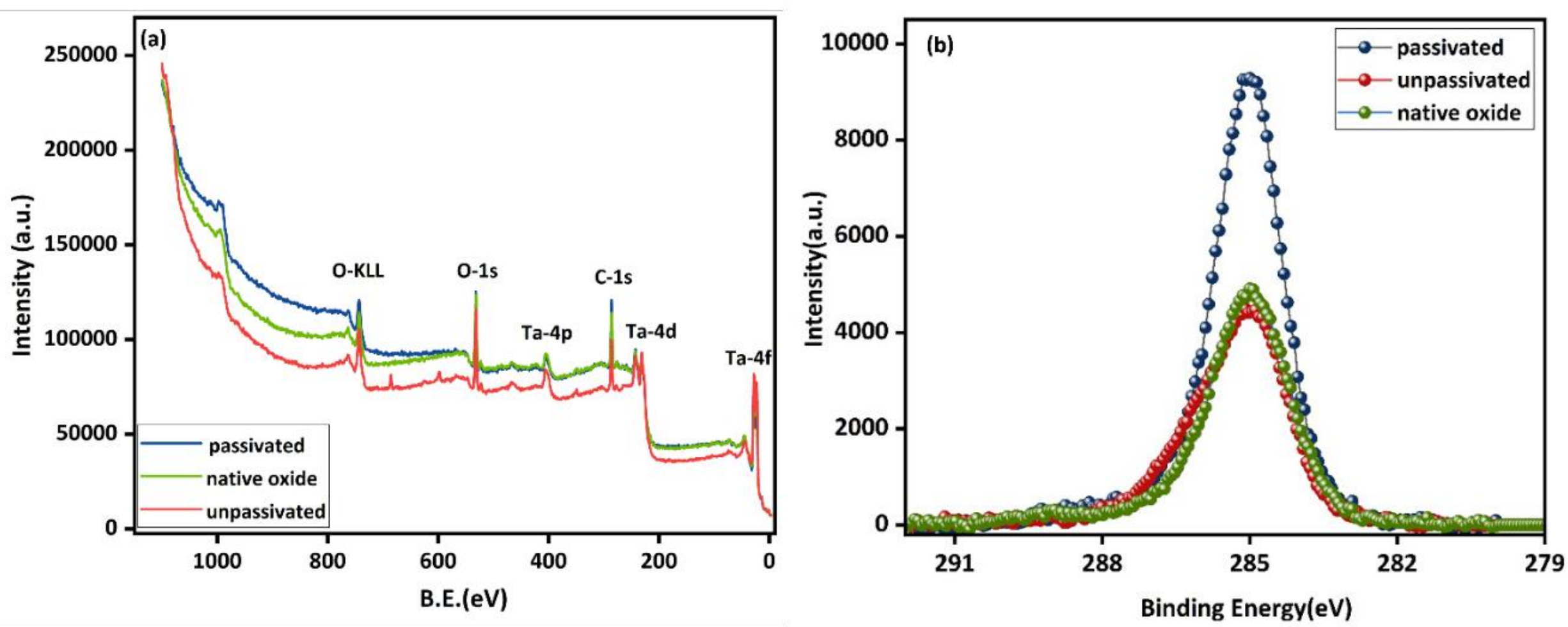

***Figure S10.*** *(a) XPS survey scan for different samples; the presence of different elements is marked. (b) Background corrected, un-normalized C 1s XPS spectra for different films. Significantly higher intensity for passivated films indicates the presence of the organic layer.*

***Table S3*** *Relative % concentration of different components for Ta 4f and C 1s XPS spectra for native oxide, unpassivated and passivated Ta thin films. Values are rounded off to closest integer.*

| | *Ta 4f* | | *C 1s* | | |
|---|---|---|---|---|---|
| | ***Ta*** | ***TaOx*** | ***C-C*** | ***C-OH/N*** | ***O=C-OH*** |
| **native oxide** | *32* | *68* | *66* | *30* | *4* |
| **unpassivated** | *42* | *58* | *60* | *37* | *3* |
| **passivated** | *51* | *49* | *81* | *13* | *7* |

*XPS Straight Line - Attenuation (SLA) Length Model*: In the SLA, it is assumed that electrons follow straight-line paths from creation to emission (i.e., no elastic scattering). In such case for an overlayer A (oxide of substrate S), the thickness ($d$) can be estimated using equation S3[8,9,10].

$$d \propto \lambda_i \, ln\left[1 + \frac{\left(\frac{I_A}{S_A}\right)}{\left(\frac{I_S}{S_S}\right)}\right] \qquad \text{(S3)}$$

Where, $\lambda_i$ is the inelastic mean free path of electrons, $I_A$ and $I_S$ are the intensities of electrons and $S_A$ and $S_S$ are the relative sensitivity factors, each for the overlayer A and the substrate S, respectively. For the estimation of thicknesses, $\lambda_i$ and sensitivity factors are extracted from the SESSA and CasaXPS software. $\lambda_i$ was calculated to be ~2 nm for tantalum oxide, the same $\lambda_i$ was assumed for all suboxides. In the case of passivated samples, an organic overlayer (thickness 1.8 nm) correction was considered to account for the damping of the Ta signal due to the SAM layer. The thickness of oxide in native oxide, unpassivated and passivated thin films was estimated using XPS to be approximately 2.5 nm, 2.1 nm and 1.4 nm, respectively. It should be noted that the oxide thickness here in case of native oxide is obtained somewhat smaller than the value estimated by ellipsometry (~3.5 nm).

*Realtive fractional contribution*: The relative fractional contribution is calculated relative to the corresponding contribution in the sample 'native oxide', and it is defined as follows:

$$\text{relative fractional contribution}\%_j = \frac{(I_{native} - I_j)}{I_{native}} \times 100$$

where $I_{native}$ is contribution of the peak under investigation in the native oxide sample and $I_j$ corresponds to the contribution of the peak under investigation in the native oxide, unpassivated or passivated sample. This results in a relative fractional contribution of '0' for the native oxide sample itself.